\documentclass[aps,prd,preprintnumbers,superscriptaddress,nofootinbib]{revtex4}
\usepackage[dvips]{graphicx}
\usepackage{bm,latexsym,amsmath,amssymb,amsfonts,mathrsfs}
\usepackage{subfigure}
\usepackage{color}
\input{colordvi.tex}
\newcommand*{\D}{{\rm d}}
\newcommand*{\mpl}{M_{\rm Pl}}
\begin{document}

\title{Full bispectra from primordial scalar and tensor perturbations in
the most general single-field inflation model}

\author{Xian~Gao}
\email[Email: ]{xgao"at"apc.univ-paris7.fr}
\affiliation{%
Astroparticule et Cosmologie (APC), UMR 7164-CNRS,
Universit\'{e} Denis Diderot-Paris 7, 10 rue Alice Domon et L\'{e}onie Duquet,
75205 Paris, France
}
\affiliation{%
Laboratoire de Physique Th\'{e}orique, \'{E}cole Normale Sup\'{e}rieure,
24 rue Lhomond, 75231 Paris, France
}
\affiliation{%
Institut d'Astrophysique de Paris (IAP), UMR 7095-CNRS,
Universit\'{e} Pierre et Marie Curie-Paris 6, 98bis Boulevard Arago, 75014 Paris, France
}

\author{Tsutomu~Kobayashi}
\email[Email: ]{tsutomu"at"rikkyo.ac.jp}
\affiliation{Department of Physics, Rikkyo University, Tokyo 171-8501, Japan}

\author{Maresuke~Shiraishi}
\email[Email: ]{mare"at"nagoya-u.jp}
\affiliation{Department of Physics and Astrophysics, Nagoya University, Aichi 464-8602, Japan}

\author{Masahide~Yamaguchi}
\email[Email: ]{gucci"at"phys.titech.ac.jp}
\affiliation{Department of Physics, Tokyo Institute of Technology, Tokyo 152-8551, Japan}

\author{Jun'ichi~Yokoyama}
\email[Email: ]{yokoyama"at"resceu.s.u-tokyo.ac.jp}
\affiliation{Research Center for the Early Universe (RESCEU), Graduate
School of Science, The University of Tokyo, Tokyo 113-0033, Japan}
\affiliation{Kavli Institute for the Physics and Mathematics of the
Universe (IPMU), The University of Tokyo, Kashiwa, Chiba, 277-8568,
Japan}

\author{Shuichiro~Yokoyama}
\email[Email: ]{shu"at"icrr.u-tokyo.ac.jp}
\affiliation{Institute for Cosmic Ray Research, The University of Tokyo,
Kashiwa, Chiba, 277-8582, Japan}

\begin{abstract}
We compute the full bispectra, namely both auto- and cross- bispectra, 
of primordial curvature and tensor perturbations in the most general 
single-field inflation model whose scalar and gravitational equations 
of motion are of second
order.
The formulae in the limits of k-inflation
and potential-driven inflation are also given. These expressions are
useful for estimating the full bispectra of temperature and polarization
anisotropies of the cosmic microwave background 
radiation.
\end{abstract}

\preprint{RUP-12-6, ICRR-Report-619-2012-08, RESCEU-32/12}
\maketitle

\section{Introduction}

The non-Gaussianities of the temperature and polarization anisotropies
of the cosmic microwave background (CMB) radiation now receive
increasing attentions because they are important tools to discriminate
models of inflation \cite{oriinf,r2}. 
Ongoing and near future project such as Planck
satellite \cite{Planck:2006aa}, CMBpol mission \cite{Baumann:2008aq},
LiteBIRD satellite \cite{LiteBIRD} would reveal the properties of the
temperature and polarization anisotropies in detail. Such 
 E-mode polarization anisotropies are sourced by both curvature
 and tensor perturbations \cite{polar}, while only tensor (and vector) perturbations can generate
B-mode polarization anisotropies \cite{Zaldarriaga:1996xe}.\footnote{Though
vector perturbations can also generate both E-mode and B-mode
polarization anisotropies, they only have a decaying mode in linear
theory and hence suppressed in the standard
inflationary cosmology based on scalar fields.} 
Therefore, even when one estimates
the ``auto'' bispectra of the temperature and the E-mode polarization
fluctuations, not only the auto bispectra but also the cross bispectra
of the primordial curvature and tensor perturbations are indispensable.

For slow-roll inflation models with the canonical kinetic term \cite{slowroll}, Maldacena
evaluated the full bispectra, including the cross bispectra, of the
primordial curvature and tensor perturbations
\cite{Maldacena:2002vr}. Inflation models are now widely generalized
into more varieties such as k-inflation \cite{ArmendarizPicon:1999rj},
DBI inflation \cite{Silverstein:2003hf}, ghost
inflation\cite{ArkaniHamed:2003uz}, G-inflation \cite{Kobayashi:2010cm},
and so on. However, almost all the works on the non-Gaussianities in
these inflation models concentrate only on the auto bispectrum of the
curvature perturbations
\cite{Seery:2005wm,Alishahiha:2004eh,Mizuno:2010ag}, which is
insufficient for evaluating the bispectra of the temperature and E-mode
polarization anisotropies of the CMB, as explained above. To our
surprise, as far as we know, the full bispectra of the primordial
curvature and tensor perturbations have not yet been obtained even for
k-inflation \cite{ArmendarizPicon:1999rj} except for
 Ref.~\cite{Shiraishi:2010kd} where  
the primordial scalar-scalar-tensor
cross bispectrum has been calculated for inflation models with
an arbitrary kinetic term.
There are several related works on the primordial cross
bispectra.  In Ref. \cite{McFadden:2011kk}, the authors show the primordial
tensor-scalar cross bispectra induced from a holographic model and
the scalar-scalar-tensor correlation has been discussed 
in the calculation of the trispectrum of the scalar fluctuations \cite{Seery:2008ax},
so-called ``graviton exchange'', and also in the context of one-loop
effects of the scalar power spectrum \cite{Bartolo:2010bu}.  
In Ref. \cite{Caldwell:2011ra},
the authors calculate the correlation between primordial scalar and vector (magnetic fields)
fluctuations in possible inflationary models of generating primordial
magnetic fields.

Among the inflation zoo, the generalized G-inflation model
\cite{Kobayashi:2011nu} occupies the unique position in that it includes
practically all the known well-behaved single inflation models since it
is based on the most general single field scalar-tensor Lagrangian with
the second order equation of motion, which was proposed by Horndeski
more than thirty years ago \cite{Horndeski} and was recently
rediscovered in the context of the generalized Galileon
\cite{Deffayet:2011gz,Charmousis:2011bf}. Indeed, it includes standard
canonical inflation \cite{oriinf,slowroll}, non-minimally coupled 
inflation \cite{nonm} including the Higgs inflation \cite{higgs},
extended inflation \cite{La:1989za}, k-inflation
\cite{ArmendarizPicon:1999rj}, DBI inflation \cite{Silverstein:2003hf},
$R^2$ inflation \cite{r2,r2m}, new Higgs inflation \cite{Germani},
G-inflation \cite{Kobayashi:2010cm}, and so on. Thus, once we analyze
properties of the primordial curvature and tensor perturbations in the
generalized G-inflation, one can apply the result for any specific
single-field inflation models. 

So far, the power spectra of scalar and tensor fluctuations were studied 
in \cite{Kobayashi:2011nu} and the general formulae for them have been
 given there.
It has been pointed out that the sound velocity
squared of the tensor perturbations as well as that of the curvature
perturbations can deviate from unity. Then the auto bispectrum of the
curvature perturbations was estimated in
Refs. \cite{Gao:2011qe,DeFelice:2011uc} 
(see also
\cite{RenauxPetel:2011sb,
Ribeiro:2012ar})
and found to be enhanced by the
inverse sound velocity squared and so on.  More recently, the auto
bispectrum of the tensor perturbations was  investigated in
Ref. \cite{gw-non-g} and found to be composed of two parts. The first
 is the universal one similar to that from Einstein gravity 
and predicts a squeezed
shape, while the other comes from the presence of the kinetic coupling
to the Einstein tensor and predicts an equilateral shape. 
What remains to be studied are the
bispectra of the primordial curvature and tensor perturbations in the
generic theory.

In the case of the most general single field model, not only auto bispectrum of
scalar perturbations but also that of tensor perturbations can be
large enough to be detected by cosmological observations, e.g., Planck satellite,
as is explained in Ref. \cite{gw-non-g}, which suggests that cross
bispectra can be large as well. For such a case, it is not necessarily
justified to consider only auto bispectrum of curvature perturbations
even when you evaluate the auto bispectrum of temperature (or E-mode)
fluctuations because cross ones can significantly contribute to it
even if the tensor-to-scalar ratio is (relatively) small. 
Furthermore, when we try to evaluate the cross bispectra including B-mode
fluctuations, the cross bispectra of tensor and scalar perturbations
are indispensable because B-mode fluctuations are produced only from tensor
perturbations. These facts are quite manifest even without any
reference nor estimation.

In such a situation, in this paper, 
we compute the cross bispectra of the primordial
curvature and tensor perturbations in the generalized G-inflation
model. The formulae in the limits of k-inflation and potential-driven
inflation are also given as specific examples.

The organization of this paper is given as follows. In the next section,
we briefly review the most general single field scalar-tensor Lagrangian
with the second order equation of motion. In Sec. III, quadratic and
cubic actions for the primordial curvature and tensor perturbations are
given. The full bispectra, including the cross ones, for them are
discussed in the section IV. The special limits for them in the cases of
k-inflation and potential driven inflation are taken in Sec. V. Final
section is devoted to conclusion and discussions.

\section{Generalized G-inflation --- The most general single-field inflation model}

The Lagrangian for the generalized G-inflation is the most general one that
is composed of the metric $g_{\mu\nu}$ and a scalar field $\phi$
together with their arbitrary derivatives but still yields the
second-order field equations.  The Lagrangian was first derived by
Horndeski in 1974 in four dimensions~\cite{Horndeski}, and very recently
it was rediscovered in a modern form as the generalized Galileon
\cite{Deffayet:2011gz}, {\em i.e.}, the most general extension of the
Galileon~\cite{Galileon, CovGali}, in arbitrary dimensions. Their
equivalence in four dimensions has been shown in Ref.~\cite{Kobayashi:2011nu}.
The four-dimensional generalized Galileon is described by the Lagrangian:
\begin{eqnarray}
\mathcal{ L}&=&K(\phi, X)-G_3(\phi, X)\Box\phi
+G_4(\phi, X)R+G_{4X}\left[(\Box\phi)^2-(\nabla_\mu\nabla_\nu\phi)^2\right]
\nonumber\\&&+
G_5(\phi, X)G_{\mu\nu}\nabla^\mu\nabla^\nu\phi
-\frac{1}{6}G_{5X}\bigl[(\Box\phi)^3
-3\Box\phi(\nabla_\mu\nabla_\nu\phi)^2+
2(\nabla_\mu\nabla_\nu\phi)^3\bigr],
\end{eqnarray}
where $K$ and $G_i$ are arbitrary functions of $\phi$ and its canonical
kinetic term
$X:=-(\partial\phi)^2/2$. We are using the notation $G_{iX}$ for
$\partial G_i/\partial X$. The generalized Galileon can be used as a
framework to study the most general single-field inflation model.
Generalized G-inflation contains novel models, as well as previously
known models of single-field inflation such as standard canonical
inflation, k-inflation, extended inflation, and new Higgs inflation, and even
$R^2$ or $f(R)$ inflation (with an appropriate field redefinition).
The above Lagrangian can also reproduce the non-minimal
coupling to the Gauss-Bonnet term~\cite{Kobayashi:2011nu}.

\section{General quadratic and cubic actions for cosmological perturbations}

In this section, we present the quadratic and cubic actions for
scalar- and tensor-type cosmological perturbations
based on the most general single-field inflation model.
Employing the Arnowitt-Deser-Misner formalism,
we write the metric as
\begin{eqnarray}
\D s^2=-N^2\D t^2+g_{ij}\left(\D x^i+N^i\D t\right)\left(\D x^j+N^j\D t\right),
\end{eqnarray}
where
\begin{eqnarray}
N=1+\alpha,\quad N_i=\partial_i\beta,
\quad
g_{ij}=a^2(t)e^{2\zeta}\left(e^h\right)_{ij},
\end{eqnarray}
and $(e^h)_{ij}=\delta_{ij}+h_{ij}+(1/2)h_{ik}h_{kj}+\cdots$.  We work
in the gauge in which the fluctuation of the scalar field vanishes,
$\phi=\phi(t)$.  Concerning the perturbations of the lapse function and
shift vector, $\alpha$ and $\beta$, it is sufficient to consider the
first order quantities to compute the cubic actions, as pointed out
in~\cite{Maldacena:2002vr}. The first order vector perturbations may be
dropped. The curvature perturbation in generalized G-inflation is shown
to be conserved on large scales at non-linear order
in~\cite{Gao:2011mz}.

Substituting the above metric to the action
and expanding it to third order,
we obtain the action for the cosmological perturbations, which will be written,
with trivial notations, as
\begin{eqnarray}
S=\int \D t\D^3x \left(
\mathcal{ L}_{hh}+\mathcal{ L}_{ss}+\mathcal{ L}_{hhh}
+\mathcal{ L}_{shh}+\mathcal{ L}_{ssh}+\mathcal{ L}_{sss}
\right).
\end{eqnarray}
The first two Lagrangians are quadratic in the metric perturbations,
which have already been obtained in Ref.~\cite{Kobayashi:2011nu}.  To define
some notations used in this paper, we will begin with summarizing the
quadratic results in the next subsection.  The third and last cubic
Lagrangians have been derived in Refs.~\cite{gw-non-g} and~\cite{Gao:2011qe,
DeFelice:2011uc}, respectively, but for completeness they are also
replicated in this section. The mixture of the scalar and tensor
perturbations, $\mathcal{ L}_{shh}$ and $\mathcal{ L}_{ssh}$, are computed for
the first time in this paper.

\subsection{Quadratic Lagrangians and primordial power spectra}

The quadratic terms are obtained as follows~\cite{Kobayashi:2011nu}.

\subsubsection{Tensor perturbations}

The most general
quadratic Lagrangian for tensor perturbations is given by
\begin{eqnarray}
\mathcal{ L}_{hh} =\frac{a^3}{8}\left[
\mathcal{ G}_T\dot h_{ij}^2-\frac{\mathcal{ F}_T}{a^2}
h_{ij,k}h_{ij,k}\right], \label{tensoraction}
\end{eqnarray}
where
\begin{eqnarray}
\mathcal{ F}_T&:=&2\left[G_4
-X\left( \ddot\phi G_{5X}+G_{5\phi}\right)\right],
\\
\mathcal{ G}_T&:=&2\left[G_4-2 XG_{4X}
-X\left(H\dot\phi G_{5X} -G_{5\phi}\right)\right].
\end{eqnarray}
Here, a dot indicates a derivative with respect to $t$, $G_{i\phi} := \partial G_i/\partial \phi$ and
the propagation speed of gravitational waves is defined as $c_h^2:=\mathcal{ F}_T/\mathcal{ G}_T$
\footnote{
In case graviton propagation speed is smaller than light speed,
nothing special happens, just the light-cone determines the causality.
In the opposite case, it has been argued that such a theory cannot
be UV completed as a Lorentz invariant theory \cite{Adams:2006sv}, though
others reach the opposite conclusions \cite{Babichev:2007dw}. We need further
investigation in this case. }
.
The linear equation of motion derived from the Lagrangian~(\ref{tensoraction})
is
\begin{eqnarray}
E^h_{ij}
:=
\partial_t\left(a^3\mathcal{ G}_T\dot h_{ij}\right)-a\mathcal{ F}_T\partial^2h_{ij} =0.
\end{eqnarray}
In deriving the above equations, we have not assumed that
the background evolution is close to de Sitter.
They can therefore be used for an arbitrary
homogeneous and isotropic
cosmological background.

We now move to the Fourier space to solve this equation:
\begin{eqnarray}
h_{ij}(t,\mathbf{x})=\int\frac{\D^3k}{(2\pi)^3}h_{ij}
(t,\mathbf{k})e^{i\mathbf{k}\cdot\mathbf{x}}.
\end{eqnarray}
It is convenient to use the conformal time coordinate defined by
$\D\eta =\D t/a$. We approximate  the inflationary regime 
 by the de Sitter spacetime and take $\mathcal{ F}_T$ and $\mathcal{ G}_T$ to be constant
\footnote{
As seen in Eqs. (\ref{eq:Fs_form}) and (\ref{eq:Gs_form}), $\mathcal{ F}_S$ and $\mathcal{ G}_S$ depend on
$\mathcal{ F}_T$ and $\mathcal{ G}_T$.  The time derivatives of $\mathcal{ F}_S$
and $\mathcal{ G}_S$ affect the spectral index of the power spectrum of
the scalar curvature perturbations and they are required to be small
from the current cosmological observations.  Hence, the assumption
that the time derivatives of $\mathcal{ F}_T$ and $\mathcal{ G}_T$ are small
are natural from observational perspectives, although one cannot rule
 out the case where $\mathcal{ F}_T$ and $\mathcal{ G}_T$ have strong 
time-dependence without conflicting
the current cosmological observations, strictly speaking.  
In this exceptional case, we must say that the assumption that the 
time derivatives of $\mathcal{ F}_T$ and $\mathcal{ G}_T$ are small is made 
just for simplicity.
}.

The quantized tensor perturbation is written as
\begin{eqnarray}
h_{ij}(\eta, \mathbf{k})=\sum_s\left[
h_{\mathbf{k}}(\eta)e_{ij}^{(s)}(\mathbf{k})a_s(\mathbf{k})
+h^*_{-\mathbf{k}}(\eta)e_{ij}^{*(s)}(-\mathbf{k})a_s^\dagger(-\mathbf{k})
\right],
\end{eqnarray}
where under these approximations the normalized mode is given by
\begin{eqnarray}
h_{\mathbf{k}}(\eta)=\frac{i\sqrt{2}H}{\sqrt{\mathcal{ F}_Tc_hk^3}}
\left(1+ic_hk\eta\right)e^{-ic_hk\eta}.
\end{eqnarray}
Here, $e_{ij}^{(s)}$ is the polarization tensor with the helicity states $s=\pm 2$,
satisfying $e_{ii}^{(s)}(\mathbf{k})=0=k_je_{ij}^{(s)}(\mathbf{k})$.
We adopt the normalization such that
\begin{eqnarray}
e_{ij}^{(s)}(\mathbf{k})e_{ij}^{*(s')}(\mathbf{k})=\delta_{ss'},
\end{eqnarray}
and choose the phase so that the following relations hold.
\begin{eqnarray}
e_{ij}^{*(s)}(\mathbf{k})=e_{ij}^{(-s)}(\mathbf{k})=e_{ij}^{(s)}(-\mathbf{k}).
\end{eqnarray}
The commutation relation for the creation and annihilation operators is
\begin{eqnarray}
[a_s(\mathbf{k}), a_{s'}^\dagger(\mathbf{k}')] =(2\pi)^3\delta_{ss'}\delta(\mathbf{k}
-\mathbf{k}').
\end{eqnarray}

The two-point function can be written as
\begin{eqnarray}
\langle h_{ij}(\mathbf{k})h_{kl}(\mathbf{k}')\rangle
&=&(2\pi)^3\delta(\mathbf{k}+\mathbf{k}')\mathcal{ P}_{ij,kl}(\mathbf{k}),
\\
\mathcal{ P}_{ij,kl}(\mathbf{k})&=&|h_{\mathbf{k}}|^2\Pi_{ij,kl}(\mathbf{k}),
\end{eqnarray}
where
\begin{eqnarray}
\Pi_{ij,kl}(\mathbf{k})=\sum_se_{ij}^{(s)}(\mathbf{k})e_{kl}^{*(s)}(\mathbf{k}).
\end{eqnarray}
The power spectrum, $\mathcal{ P}_h=(k^3/2\pi^2)\mathcal{ P}_{ij,ij}$, is
thus computed as
\begin{eqnarray}
\mathcal{ P}_h=\left.\frac{2}{\pi^2}\frac{H^2}{\mathcal{ F}_Tc_h}\right|_{c_hk\eta=-1}.
\label{eq:tensorpower}
\end{eqnarray}

\subsubsection{Scalar perturbations}

The quadratic Lagrangian for the scalar perturbations is given by
\begin{eqnarray}
\mathcal{ L}_{ss}= a^3\left[
-3\mathcal{ G}_{T}\dot\zeta^2+\frac{\mathcal{ F}_T}{a^2}\zeta_{,i}\zeta_{,i}
+\Sigma \alpha^2
-\frac{2}{a^2}\Theta\alpha\beta_{,ii}
+\frac{2}{a^2}\mathcal{ G}_T\dot \zeta\beta_{,ii}
+6\Theta \alpha\dot\zeta
-\frac{2}{a^2}\mathcal{ G}_T\alpha\zeta_{,ii}
\right],\label{scalaraction1}
\end{eqnarray}
where
\begin{eqnarray}
\Sigma&:=&XK_X+2X^2K_{XX}+12H\dot\phi XG_{3X}
+6H\dot\phi X^2G_{3XX}
-2XG_{3\phi}-2X^2G_{3\phi X}
\nonumber\\&&
-6H^2G_4
+6\left[H^2\left(7XG_{4X}+16X^2G_{4XX}+4X^3G_{4XXX}\right)
\right.
\nonumber\\&&
\left.
-H\dot\phi\left(G_{4\phi}+5XG_{4\phi X}+2X^2G_{4\phi XX}\right)
\right]
\nonumber\\&&
+30H^3\dot\phi XG_{5X}+26H^3\dot\phi X^2G_{5XX}
+4H^3\dot\phi X^3G_{5XXX} 
\nonumber\\&&
-6H^2X\left(6G_{5\phi}
+9XG_{5\phi X}+2 X^2G_{5\phi XX}\right),
\\
\Theta&:=&-\dot\phi XG_{3X}+
2HG_4-8HXG_{4X}
-8HX^2G_{4XX}+\dot\phi G_{4\phi}+2X\dot\phi G_{4\phi X}
\nonumber\\&&
-H^2\dot\phi\left(5XG_{5X}+2X^2G_{5XX}\right)
+2HX\left(3G_{5\phi}+2XG_{5\phi X}\right).
\end{eqnarray}

Varying Eq.~(\ref{scalaraction1}) with respect to $\alpha$ and $\beta$,
we get the first-order constraint equations:
\begin{eqnarray}
\Sigma\alpha -\frac{\Theta}{a^2}\partial^2\beta +3\Theta\dot\zeta
-\frac{\mathcal{ G}_T}{a^2}\partial^2\zeta&=&0,
\\
\Theta\alpha-\mathcal{ G}_T\dot\zeta&=&0,
\end{eqnarray}
which are solved to yield
\begin{eqnarray}
\alpha&=&\frac{\mathcal{ G}_T}{\Theta}\dot\zeta,\label{csol1}
\\
\beta
&=&\frac{1}{a\mathcal{ G}_T}\left(a^3\mathcal{ G}_S\psi
-\frac{a\mathcal{ G}_T^2}{\Theta}\zeta\right),\label{csol2}
\end{eqnarray}
with $\psi:=\partial^{-2}\dot\zeta$.  Plugging Eqs.~(\ref{csol1})
and~(\ref{csol2}) to Eq.~(\ref{scalaraction1}), we obtain
\begin{eqnarray}
\mathcal{ L}_{ss}=a^3\left[
\mathcal{ G}_S
\dot\zeta^2
-\frac{\mathcal{ F}_S}{a^2}
\zeta_{,i}\zeta_{,i}
\right]\label{scalar2},
\end{eqnarray}
where we have defined
\begin{eqnarray}
\mathcal{ F}_S&:=&\frac{1}{a}\frac{\D}{\D t}\left(\frac{a}{\Theta}\mathcal{ G}_T^2\right)
-\mathcal{ F}_T, \label{eq:Fs_form}
\\
\mathcal{ G}_S&:=&\frac{\Sigma }{\Theta^2}\mathcal{ G}_T^2+3\mathcal{ G}_T. \label{eq:Gs_form}
\end{eqnarray}
The sound speed is given by $c_s^2:=\mathcal{ F}_S/\mathcal{ G}_S$.
The linear equation of motion derived
from the Lagrangian~(\ref{scalar2}) is
\begin{eqnarray}
E^s
:=
\partial_t\left(a^3\mathcal{ G}_S\dot\zeta\right)-a\mathcal{ F}_S\partial^2\zeta =0.
\end{eqnarray}

The scalar two-point function can be calculated in a way similar to the
case of the tensor perturbations.  We move to the Fourier space:
\begin{eqnarray}
\zeta(t,\mathbf{x})=\int\frac{\D^3k}{(2\pi)^3}\zeta
(t,\mathbf{k})e^{i\mathbf{k}\cdot\mathbf{x}},
\end{eqnarray}
and proceed in the de Sitter approximation, assuming that $\mathcal{ F}_S$
and $\mathcal{ G}_S$ are almost constant.
The quantized curvature
perturbation is written as
\begin{eqnarray}
\zeta(\eta, \mathbf{k})=
\xi_{\mathbf{k}}(\eta)a(\mathbf{k})
+\xi^*_{-\mathbf{k}}(\eta)a^\dagger(-\mathbf{k}),
\end{eqnarray}
where the normalized mode is given by
\begin{eqnarray}
\xi_{\mathbf{k}}(\eta)=\frac{iH}{2\sqrt{\mathcal{ F}_Sc_sk^3}}
\left(1+ic_sk\eta\right)e^{-c_sk\eta}.
\end{eqnarray}
The commutation relation for the creation and annihilation operators is
\begin{eqnarray}
[a(\mathbf{k}), a^\dagger(\mathbf{k}')] =(2\pi)^3\delta(\mathbf{k}
-\mathbf{k}').
\end{eqnarray}

Thus, the power spectrum is calculated as
\begin{eqnarray}
\langle\zeta(\mathbf{k})\zeta(\mathbf{k}')\rangle
&=&(2\pi)^3\delta(\mathbf{k}+\mathbf{k}')\frac{2\pi^2}{k^3}\mathcal{ P}_\zeta,
\\
\mathcal{ P}_\zeta&=&\left.\frac{1}{8\pi^2}\frac{H^2}{\mathcal{ F}_Sc_s}\right|_{c_sk\eta=-1}.
\label{eq:scalarpower}
\end{eqnarray}
From Eqs. (\ref{eq:tensorpower}) and (\ref{eq:scalarpower}),
tensor-to-scalar ratio $r$ is given by
\begin{eqnarray}
r := \frac{\mathcal{ P}_h}{\mathcal{ P}_\zeta} = 16 \frac{\mathcal{ F}_S c_s}{\mathcal{
 F}_T c_h}~, \label{tensortoscalar}
\end{eqnarray}
where we have assumed that the relevant quantities remain practically
constant between the horizon crossings of tensor and scalar
perturbations that occur at different time in case
 $c_h \neq c_s$ \cite{Lorenz:2008et}.

\subsection{Cubic Lagrangians}

We now present the most general cubic Lagrangians
composed of the tensor and scalar perturbations.
We would like to emphasize that
in deriving the following Lagrangians
the slow-roll approximation is {\em not} used,
as discussed in literature \cite{Khoury:2008wj}.

\subsubsection{Three tensors}

The Lagrangian involving three tensors was derived in Ref.~\cite{gw-non-g}:
\begin{eqnarray}
\mathcal{ L}_{hhh}=
a^3\left[
\frac{\mu }{12} \dot h_{ij}\dot h_{jk}\dot h_{ki}
+\frac{\mathcal{ F}_T}{4a^2}\left(h_{ik}h_{jl}
-\frac{1}{2}h_{ij}h_{kl}\right)h_{ij,kl}
\right],
\end{eqnarray}
where we defined
\begin{eqnarray}
\mu :=\dot\phi XG_{5X}.\label{mudefine}
\end{eqnarray}
As discussed in Ref.~\cite{gw-non-g}, this cubic action for the
tensor perturbation $h_{ij}$ is composed only of two contributions. The
former has one time derivative on each $h_{ij}$ and newly appears in the
presence of the kinetic coupling to the Einstein tensor, that is,
$G_{5X} \ne 0$. On the other hand, the latter has two spacial
derivatives and is essentially identical to the cubic term that appears
in Einstein gravity. Therefore, in what follows, we use the
terminologies "new" and "GR" for corresponding terms.

\subsubsection{Two tensors and one scalar}

The interactions involving two tensors and one scalar
are given by
\begin{eqnarray}
\mathcal{ L}_{shh}&=&a^3\left[
\frac{3\mathcal{ G}_T}{8}\zeta\dot h_{ij}^2-\frac{\mathcal{ F}_T}{8a^2}\zeta
h_{ij,k}h_{ij,k}-\frac{\mu}{4}\dot\zeta\dot h_{ij}^2
-\frac{\Gamma}{8}\alpha \dot h_{ij}^2-\frac{\mathcal{ G}_T}{8a^2}\alpha h_{ij,k}h_{ij,k}
-\frac{\mu}{2a^2}\alpha\dot h_{ij}h_{ij,kk}
\right]
\nonumber\\&&
-a\left[\frac{\mathcal{ G}_T}{4}\beta_{,k}\dot h_{ij}h_{ij,k}
+\frac{\mu }{2}
\left(\dot h_{ik}\dot h_{jk}\beta_{,ij}-\frac{1}{2}
\dot h_{ij}^2\beta_{,kk}\right)
\right],\label{tts1}
\end{eqnarray}
where
\begin{eqnarray}
\Gamma&:=& 2G_4-8XG_{4X}-8X^2G_{4XX}
\nonumber\\&&
-2H\dot\phi\left(5XG_{5X}+2X^2G_{5XX}\right)
+2X\left(3G_{5\phi}+2XG_{5\phi X}\right).
\end{eqnarray}
This quantity can also be expressed in a compact form
 $\Gamma=\partial\Theta/\partial H$.

Substituting the first-order constraint equations
to Eq.~(\ref{tts1}), the Lagrangian reduces to
\begin{eqnarray}
\mathcal{ L}_{shh}&=&a^3\left[
b_1\zeta\dot h_{ij}^2+\frac{b_2}{a^2}\zeta h_{ij,k}h_{ij,k}
+b_3\psi_{,k}\dot h_{ij}h_{ij,k}+b_4\dot\zeta\dot h_{ij}^2
+\frac{b_5}{a^2}\partial^2\zeta \dot h_{ij}^2\right.
\nonumber\\&&
\left.
+b_6\psi_{,ij}\dot h_{ik}\dot h_{jk}
+\frac{b_7}{a^2}\zeta_{,ij}\dot h_{ik}\dot h_{jk}\right]+E_{shh},
\end{eqnarray}
where
\begin{eqnarray}
b_1&=&\frac{3\mathcal{ G}_T}{8}\left[1-\frac{H\mathcal{ G}_T^2}{\Theta\mathcal{ F}_T}
+\frac{\mathcal{ G}_T}{3}\frac{\D}{\D t}\left(\frac{\mathcal{ G}_T}{\Theta\mathcal{ F}_T}\right)
\right],
\\
b_2&=&\frac{\mathcal{ F}_S}{8},
\\
b_3&=&-\frac{\mathcal{ G}_S}{4},
\\
b_4&=&\frac{\mathcal{ G}_T}{8\Theta \mathcal{ F}_T}\left(\mathcal{ G}_T^2-\Gamma\mathcal{ F}_T\right)
+\frac{\mu}{4}\left[
\frac{\mathcal{ G}_S}{\mathcal{ G}_T}-1-\frac{H\mathcal{ G}_T^2}{\Theta\mathcal{ F}_T}
\left(6+\frac{\dot{\mathcal{ G}}_S}{H\mathcal{ G}_S}\right)\right]
+\frac{\mathcal{ G}_T^2}{4}\frac{\D}{\D t}\left(\frac{\mu}{\Theta\mathcal{ F}_T}\right),
\\
b_5&=&\frac{\mu\mathcal{ G}_T}{4\Theta}\left(\frac{\mathcal{ F}_S\mathcal{ G}_T}{\mathcal{ F}_T\mathcal{ G}_S}
-1\right),
\\
b_6&=&-\frac{\mu}{2}\frac{\mathcal{ G}_S}{\mathcal{ G}_T},
\\
b_7&=&\frac{\mu}{2}\frac{\mathcal{ G}_T}{\Theta},
\end{eqnarray}
and
\begin{eqnarray}
E_{shh}=\frac{\mu}{4\mathcal{ G}_S}\frac{\mathcal{ G}_T^2}{\Theta\mathcal{ F}_T}\dot h_{ij}^2E^s
+\frac{\mathcal{ G}_T^2}{2\Theta\mathcal{ F}_T}
\left(\frac{\zeta}{2}+\frac{\mu}{\mathcal{ G}_T}\dot\zeta\right)\dot h_{ij}E_{ij}^h.
\end{eqnarray}
The last term $E_{shh}$ can be removed by redefining the fields as
\begin{eqnarray}
h_{ij}&\to&h_{ij}+\frac{\mathcal{ G}_T^2}{\Theta\mathcal{ F}_T}
\left(\zeta+\frac{2\mu}{\mathcal{ G}_T}\dot\zeta\right)\dot h_{ij},
\\
\zeta&\to&\zeta + \frac{\mu}{8\mathcal{ G}_S}\frac{\mathcal{ G}_T^2}{\Theta\mathcal{ F}_T}\dot h_{ij}^2.
\end{eqnarray}
The contribution to the correlation function
is however negligible because the above field redefinitions involve
at least one time derivative of the metric perturbation,
which vanishes on super-horizon scales.

\subsubsection{Two scalars and one tensor}

The interactions involving one tensor and two scalars
are given by
\begin{eqnarray}
\mathcal{ L}_{ssh}&=&a\left[
2\Theta \alpha\beta_{,ij}h_{ij} +\frac{\Gamma }{2}
\alpha\beta_{,ij}\dot h_{ij}
+\frac{\mu}{a^2}\alpha\beta_{,ij}h_{ij,kk}
-\frac{3\mathcal{ G}_T}{2}\zeta\beta_{,ij}\dot h_{ij}
-2\mathcal{ G}_T\dot\zeta\beta_{,ij}h_{ij}
+\mu \dot\zeta\beta_{,ij}\dot h_{ij}\right.
\nonumber\\&&
\left.
-\mathcal{ F}_T\zeta_{,i}\zeta_{,j}h_{ij}-2\mathcal{ G}_T\alpha_{,i}\zeta_{,j}h_{ij}
+\mu\alpha_{,i}\zeta_{,j}\dot h_{ij}
+\frac{\mathcal{ G}_T}{2a^2}\beta_{,ij}\beta_{,k}h_{ij,k}
+\frac{\mu }{a^2}\beta_{,ij}\beta_{,k}\dot h_{ij,k}
\right].
\end{eqnarray}

Substituting the constraint equations, we obtain the reduced Lagrangian:
\begin{eqnarray}
\mathcal{ L}_{ssh}&=&a^3\left[\frac{c_1}{a^2}
h_{ij}\zeta_{,i}\zeta_{,j}
+\frac{c_2}{a^2}\dot h_{ij}\zeta_{,i}\zeta_{,j}
+c_3\dot h_{ij}\zeta_{,i}\psi_{,j}
+\frac{c_4}{a^2}\partial^2h_{ij}\zeta_{,i}\psi_{,j}\right.
\nonumber\\&&
\left.
+\frac{c_5}{a^4}\partial^2 h_{ij}\zeta_{,i}\zeta_{,j}
+c_6\partial^2 h_{ij}\psi_{,i}\psi_{,j}\right]+E_{ssh}, \label{ssh}
\end{eqnarray}
where
\begin{eqnarray}
c_1&=&\mathcal{ F}_S,
\\
c_2&=&\frac{\Gamma}{4\Theta}\left(\mathcal{ F}_S-\mathcal{ F}_T\right)
+\frac{\mathcal{ G}_T^2}{\Theta}\left[
-\frac{1}{2}+\frac{H\Gamma}{4\Theta}\left(3+\frac{\dot{ \mathcal{ G}}_T}{H\mathcal{ G}_T}\right)
-\frac{1}{4}\frac{\D}{\D t}\left(\frac{\Gamma}{\Theta}\right)
\right]
\nonumber\\&&
+\frac{\mu\mathcal{ F}_S}{\mathcal{ G}_T}+\frac{2H\mathcal{ G}_T\mu}{\Theta}
-\mathcal{ G}_T\frac{\D}{\D  t}\left(\frac{\mu}{\Theta}\right),
\\
c_3&=&\mathcal{ G}_S\left[
\frac{3}{2}+\frac{\D}{\D t}\left(\frac{\Gamma}{2\Theta}+\frac{\mu}{\mathcal{ G}_T}\right)
-\left(3H+\frac{\dot{\mathcal{ G}}_T}{\mathcal{ G}_T}\right)
\left(\frac{\Gamma}{2\Theta}+\frac{\mu}{\mathcal{ G}_T}\right)
\right],
\\
c_4&=&\mathcal{ G}_S\left[
-\frac{\mathcal{ G}_T^2-\Gamma\mathcal{ F}_T}{2\Theta\mathcal{ G}_T}
-\frac{2H\mu}{\Theta}+\frac{\D}{\D t}\left(\frac{\mu}{\Theta}\right)
+\frac{\mu}{\mathcal{ G}_T^2}\left(\mathcal{ F}_T-\mathcal{ F}_S\right)
\right],
\\
c_5&=&\frac{\mathcal{ G}_T^2}{2\Theta}\left[
\frac{\mathcal{ G}_T^2-\Gamma\mathcal{ F}_T}{2\Theta\mathcal{ G}_T}
+\frac{2H\mu}{\Theta}-\frac{\D}{\D t}\left(\frac{\mu}{\Theta}\right)
-\frac{\mu}{\mathcal{ G}_T^2}\left(3\mathcal{ F}_T-\mathcal{ F}_S\right)
\right],
\\
c_6&=&\frac{\mathcal{ G}_S^2}{4\mathcal{ G}_T}\left[
1+\frac{6H\mu}{\mathcal{ G}_T}-2\mathcal{ G}_T\frac{\D}{\D t}\left(\frac{\mu}{\mathcal{ G}_T^2}\right)
\right],
\end{eqnarray}
and
\begin{eqnarray}
E_{ssh}=\bar f_i\partial^{-2}\partial_iE^s+\bar f_{ij}E_{ij}^h,
\end{eqnarray}
with
\begin{eqnarray}
\bar f_i&:=&\frac{\Gamma}{2\Theta}\zeta_{,j}h_{ij}
+\frac{\mu}{\mathcal{ G}_T}\zeta_{,j}\dot h_{ij}
+\frac{\mu}{a^2\Theta}\zeta_{,j}\partial^2 h_{ij}
-\frac{\mu\mathcal{ G}_S}{\mathcal{ G}_T^2}\psi_{,j}\partial^2 h_{ij},
\\
\bar f_{ij}&:=&\frac{\mathcal{ G}_S}{\Theta\mathcal{ G}_T}\left(\frac{\Gamma}{2}
+\frac{\mu\Theta}{\mathcal{ G}_T}\right)\zeta_{,i}\psi_{,j}
-\frac{\mathcal{ G}_T}{a^2\Theta^2}\left(\frac{\Gamma}{4}+\frac{\mu \Theta}{\mathcal{ G}_T}\right)
\zeta_{,i}\zeta_{,j}.
\end{eqnarray}

The field redefinition:
\begin{eqnarray}
h_{ij}&\to& h_{ij}+4\bar f_{ij},
\\
\zeta&\to&\zeta-\frac{1}{2}\partial^{-2}\partial_i\bar f_{i},
\end{eqnarray}
removes the last term $E_{ssh}$.
Since all the terms involve at least one derivative of the metric perturbation,
the field redefinition does not contribute to the correlation function
on super-horizon scales.

\subsubsection{Three scalars}

For completeness, here we give the cubic Lagrangian for the scalar
perturbations derived in Refs.~\cite{Gao:2011qe, DeFelice:2011uc}. The
cubic Lagrangian for the scalar perturbations is given by
\begin{eqnarray}
\mathcal{ L}_{sss}&=&
-\frac{a^3}{3}\left(\Sigma+2X\Sigma_X+H\Xi\right)\alpha^3
+a^3\left[3\Sigma\zeta+\Xi\dot\zeta+\left(\Gamma-\mathcal{ G}_T\right)
\frac{\zeta_{,ii}}{a^2}-\frac{\Xi}{3a^2}\beta_{,ii}
\right]\alpha^2
\nonumber\\&&
-2a\Theta\alpha\zeta_{,i}\beta_{,i}+18a^3\Theta\alpha\zeta\dot\zeta
+4a\mu\alpha\dot\zeta\zeta_{,ii}
-\frac{\Gamma}{2a}\alpha\left(\beta_{,ij}\beta_{,ij}-\beta_{,ii}\beta_{,jj}\right)
\nonumber\\&&
+\frac{2\mu}{a}\alpha\left(\beta_{,ij}\zeta_{,ij}-\beta_{,ii}\zeta_{,jj}\right)
-2a\Theta\alpha\beta_{,ii}\zeta - 2a\Gamma\alpha\beta_{,ii}\dot\zeta
-2a\mathcal{ G}_T\alpha\zeta\zeta_{,ii}
-a\mathcal{ G}_T\alpha\zeta_{,i}\zeta_{,i}
\nonumber\\&&
+
3a^3\Gamma\alpha\dot\zeta^2
+2a^3\mu\dot\zeta^3+a\mathcal{ F}_T\zeta\zeta_{,i}\zeta_{,i }
-9a^3\mathcal{ G}_T\dot\zeta^2\zeta+2a\mathcal{ G}_T\beta_{,i}\zeta_{,i}\dot\zeta
-2a\mu\beta_{ii}\dot\zeta^2
\nonumber\\&&
+2a\mathcal{ G}_T\beta_{,ii}\dot\zeta\zeta
+\frac{1}{a}\left(\frac{3}{2}\mathcal{ G}_T\zeta -\mu\dot\zeta\right)
\left(\beta_{,ij}\beta_{,ij}-\beta_{,ii}\beta_{,jj}\right)
-2\frac{\mathcal{ G}_T}{a}\beta_{,ii}\beta_{,j}\zeta_{,j},
\end{eqnarray}
where
\begin{eqnarray}
\Xi &:=&
12\dot\phi XG_{3X}
+6\dot\phi X^2G_{3XX}
-12HG_4
\nonumber\\&&
+6\left[2H\left(7XG_{4X}+16X^2G_{4XX}+4X^3G_{4XXX}\right)
-\dot\phi\left(G_{4\phi}+5XG_{4\phi X}+2X^2G_{4\phi XX}\right)
\right]
\nonumber\\&&
+90H^2\dot\phi XG_{5X}+78H^2\dot\phi X^2G_{5XX}
+12H^2\dot\phi X^3G_{5XXX} \nonumber\\&&
-12HX\left(6G_{5\phi}
+9XG_{5\phi X}+2 X^2G_{5\phi XX}\right).
\end{eqnarray}
Using the first-order constraint equations to remove $\alpha$ and $\beta$
from the above Lagrangian, we obtain the following reduced expression:
\begin{eqnarray}
\mathcal{ L}_{sss}= \int dtd^{3}xa^{3}\mathcal{G}_{S}\left[\frac{\mathcal{C}_{1}}{6H}\dot{\zeta}^{3}+\mathcal{C}_{2}\dot{\zeta}^{2}\zeta+\mathcal{C}_{3}\frac{2c_{s}^{2}}{a^{2}}\zeta\left(\partial_{i}\zeta\right)^{2}+2\mathcal{C}_{4}\dot{\zeta}\partial_{i}\zeta\partial^{i}\psi+2\mathcal{C}_{5}\partial^{2}\zeta\left(\partial_{i}\psi\right)^{2}\right],
\end{eqnarray}
with $\psi = \partial^{-2}\dot{\zeta}$.
There are five independent cubic terms with coefficients:
\begin{eqnarray}
\mathcal{C}_{1} & = & -\frac{8\Xi\mathcal{G}_{T}^{3}}{3\Theta^{3}\mathcal{G}_{S}}+\frac{2H^{2}}{\Theta\mathcal{F}_{S}}\left[\frac{2\Xi\mathcal{G}_{T}^{3}}{\Theta^{2}}+\frac{3\mathcal{G}_{T}^{3}}{\Theta\mathcal{F}_{S}}\left(\mathcal{G}_{S}-2\mathcal{F}_{S}\right)+36\mu\left(\mathcal{G}_{T}-\mathcal{G}_{S}\right)+\frac{9\Gamma}{\Theta}\mathcal{G}_{T}\left(2\mathcal{G}_{T}-\mathcal{G}_{S}\right)\right]\nonumber \\
 &  & +2H\left[6\mu\left(\frac{1}{\mathcal{G}_{S}}-\frac{1}{\mathcal{G}_{T}}\right)+\frac{2\left(\Sigma-X\Sigma_{X}\right)\mathcal{G}_{T}^{3}}{\Theta^{3}\mathcal{G}_{S}}+\frac{\Xi\mathcal{G}_{T}}{\Theta^{2}}\left(\frac{3\mathcal{G}_{T}}{\mathcal{G}_{S}}-1\right)\right.\nonumber \\
 &  & \qquad\left.+\frac{3\mathcal{G}_{T}}{\Theta}\left(\frac{\mathcal{G}_{S}}{\mathcal{F}_{S}}+\frac{3\mathcal{G}_{T}}{\mathcal{G}_{S}}-1\right)+3\frac{\Gamma}{\Theta}\left(\frac{3\mathcal{G}_{T}}{\mathcal{G}_{S}}-2\right)\right]-\frac{6H^{3}\mathcal{G}_{S}\mathcal{G}_{T}^{2}}{\Theta^{2}\mathcal{F}_{S}^{2}}\left(6\mu+\frac{\Gamma\mathcal{G}_{T}}{\Theta}\right),\label{C1}
\end{eqnarray}
\begin{eqnarray}
\mathcal{C}_{2}&=&3+3H\mathcal{G}_{S}\left(\frac{\mu}{\mathcal{G}_{T}^{2}}+\frac{\Gamma}{2\Theta\mathcal{G}_{T}}-\frac{3\mathcal{G}_{T}}{2\Theta\mathcal{F}_{S}}\right)
\nonumber\\&&
+3\frac{H^{2}\mathcal{G}_{S}}{\Theta\mathcal{F}_{S}}\left(8\mu+\frac{2\Gamma\mathcal{G}_{T}}{\Theta}-\frac{\mathcal{G}_{T}^{3}}{2\Theta\mathcal{F}_{S}}\right)+\frac{3H^{3}\mathcal{G}_{S}\mathcal{G}_{T}^{2}}{\Theta^{2}\mathcal{F}_{S}^{2}}\left(3\mu+\frac{\Gamma\mathcal{G}_{T}}{2\Theta}\right),\label{C2}
\end{eqnarray}
\begin{eqnarray}
\mathcal{C}_{3}&=&\frac{\mathcal{F}_{T}}{2\mathcal{F}_{S}}+H\left[\frac{\left(3\mathcal{G}_{S}-2\mathcal{G}_{T}\right)\mathcal{G}_{T}}{4\Theta\mathcal{F}_{S}}-\frac{\mu\mathcal{G}_{S}}{2\mathcal{G}_{T}^{2}}-\frac{\Gamma\mathcal{G}_{S}}{4\Theta\mathcal{G}_{T}}\right]
\nonumber\\&&
+\frac{H^{2}\mathcal{G}_{S}}{\Theta\mathcal{F}_{S}}\left(\frac{\mathcal{G}_{T}^{3}}{4\Theta\mathcal{F}_{S}}-4\mu-\frac{\Gamma\mathcal{G}_{T}}{\Theta}\right)-\frac{H^{3}\mathcal{G}_{S}\mathcal{G}_{T}^{2}}{2\Theta^{2}\mathcal{F}_{S}^{2}}\left(3\mu+\frac{\Gamma\mathcal{G}_{T}}{2\Theta}\right),\label{C3}
\end{eqnarray}
\begin{equation}
\mathcal{C}_{4}=-\frac{\mathcal{G}_{S}}{4\mathcal{G}_{T}}+3H\mathcal{G}_{S}\left(\frac{\mu}{2\mathcal{G}_{T}^{2}}+\frac{\Gamma}{4\Theta\mathcal{G}_{T}}-\frac{\mathcal{G}_{T}}{2\Theta\mathcal{F}_{S}}\right)+3\frac{H^{2}\mathcal{G}_{S}}{\Theta\mathcal{F}_{S}}\left(2\mu+\frac{\Gamma\mathcal{G}_{T}}{2\Theta}\right),\label{C4}
\end{equation}
\begin{equation}
\mathcal{C}_{5}=\frac{3\mathcal{G}_{S}}{8\mathcal{G}_{T}}-\frac{3H\mathcal{G}_{S}}{4\mathcal{G}_{T}}\left(\frac{\mu}{\mathcal{G}_{T}}+\frac{\Gamma}{2\Theta}\right).\label{C5}
\end{equation}

\section{Primordial bispectra}

Having obtained the general cubic Lagrangians
composed of the scalar and tensor perturbations,
we now compute the bispectra in this section.
Here, we use the mode functions in exact de Sitter.

\subsection{Three tensors}

Let us consider three-point function of the tensor perturbations:
\begin{eqnarray}
&&\langle h_{i_1j_1}(\mathbf{k}_1)
h_{i_2j_2}(\mathbf{k}_2)
h_{i_3j_3}(\mathbf{k}_3)
\rangle
 = (2\pi)^3\delta(\mathbf{k}_1+\mathbf{k}_2+\mathbf{k}_3)
B_{i_1j_1i_2j_2i_3j_3}^{(hhh)},
\\
&&
B_{i_1j_1i_2j_2i_3j_3}^{(hhh)} =
\frac{(2\pi)^4\mathcal{ P}_h^2}{k_1^3k_2^3k_3^3}
\left(
{\widetilde{\mathcal{ A}}}^{({\rm new})}_{i_1j_1i_2j_2i_3j_3}+{\widetilde{\mathcal{ A}}}^{({\rm GR})}_{i_1j_1i_2j_2i_3j_3}
\right), \label{eq:Bhhh}
\end{eqnarray}
where ${\widetilde{\mathcal{ A}}}^{({\rm new})}_{i_1j_1i_2j_2i_3j_3}$ and
${\widetilde{\mathcal{ A}}}^{({\rm GR})}_{i_1j_1i_2j_2i_3j_3}$ represent the
contributions from the $\dot h^3$ term and the $h^2\partial^2 h$ terms,
respectively. 

Each contribution is given by
\begin{eqnarray}
{\widetilde{\mathcal{ A}}}_{i_1j_1i_2j_2i_3j_3}^{({\rm new})} &=&
\frac{H\mu}{4\mathcal{ G}_T}
\frac{k_1^2k_2^2k_3^2}{K^3}\Pi_{i_1j_1, lm}(\mathbf{k}_1)
\Pi_{i_2j_2, mn}(\mathbf{k}_2)
\Pi_{i_3j_3, nl}(\mathbf{k}_3), \label{eq:Anewtype}
\\
{\widetilde{\mathcal{ A}}}_{i_1j_1i_2j_2i_3j_3}^{({\rm GR})} &=&
\widetilde{\mathcal{ A}}
\left\{
\Pi_{i_1j_1,ik}(\mathbf{k}_1)\Pi_{i_2j_2,jl}(\mathbf{k}_2)
\left[
k_{3k}k_{3l}\Pi_{i_3j_3,ij}(\mathbf{k}_3)
-\frac{1}{2}k_{3i}k_{3k}\Pi_{i_3j_3,jl}(\mathbf{k}_3)
\right] \right.
\nonumber\\&&
\left.
+5~{\rm perms}~{\rm of}~ 1, 2, 3
\right\},\;\;\; \label{eq:AGRtype}
\end{eqnarray}
where $K=k_1+k_2+k_3$ and
\begin{eqnarray}
\widetilde{\mathcal{ A}}(k_1,k_2,k_3):=
-\frac{K}{16}\biggl[
1-\frac{1}{K^3}\sum_{i\neq j}k_i^2k_j-4\frac{k_1k_2k_3}{K^3}
\biggr].\;\;\;
\end{eqnarray}
The first term ${\widetilde{\mathcal{ A}}}_{i_1j_1i_2j_2i_3j_3}^{({\rm new})}$
is proportional to $G_{5X}$ and hence vanishes in the case of Einstein gravity,
while the second term ${\widetilde{\mathcal{ A}}}_{i_1j_1i_2j_2i_3j_3}^{({\rm GR})}$
is universal in the sense that
it is independent of any model parameters
and remains the same even in non-Einstein gravity.

In order to quantify the magnitude of the bispectrum, we define two
polarization modes as
\begin{equation}
  \xi^{(s)}(\mathbf{k}) := h_{ij}(\mathbf{k})e^{*(s)}_{ij}(\mathbf{k}),
\end{equation}
and their relevant amplitudes of the bispectra as
\begin{equation}
  \langle \xi^{(s_1)}(\mathbf{k}_1)
\xi^{(s_2)}(\mathbf{k}_2)
\xi^{(s_3)}(\mathbf{k}_3)
\rangle
=(2\pi)^7\delta(\mathbf{k}_1+\mathbf{k}_2+\mathbf{k}_3)
\frac{\mathcal{ P}_h^2}{k_1^3k_2^3k_3^3}
\left(
{\widetilde{\mathcal{ A}}}_{({\rm new})}^{s_1 s_2 s_3}+{\widetilde{\mathcal{ A}}}_{({\rm GR})}^{s_1 s_2 s_3}
\right).
\end{equation}
From Eqs.~(\ref{eq:Anewtype}) and~(\ref{eq:AGRtype}), the amplitudes
$\widetilde{\mathcal{ A}}_{({\rm new}),({\rm GR})}^{s_1 s_2 s_3}$ are easily
calculated as \cite{gw-non-g}
\begin{eqnarray}
 {\widetilde{\mathcal{ A}}}_{({\rm new})}^{s_1 s_2 s_3} &=&
 \frac{H\mu}{4\mathcal{ G}_T}\frac{k_1^2k_2^2k_3^2}{K^3}F(s_1k_1, s_2k_2, s_3k_3),
 \\
  {\widetilde{\mathcal{ A}}}_{({\rm GR})}^{s_1 s_2 s_3} &=&
  \frac{\widetilde{\mathcal{ A}}}{2}\left(s_1k_1+s_2k_2+s_3k_3\right)^2
  F(s_1k_1, s_2k_2, s_3k_3),
\end{eqnarray}
where
\begin{eqnarray}
F(x,y,z):=\frac{1}{64}\frac{1}{x^2y^2z^2}
(x+y+z)^3(x-y+z)(x+y-z)(x-y-z).
\end{eqnarray}
As pointed out in Ref.~\cite{gw-non-g},
${\widetilde{\mathcal{ A}}}_{({\rm new})}^{+++}$ has a peak in the equilateral limit, while
${\widetilde{\mathcal{ A}}}_{({\rm GR})}^{+++}$ in the squeezed limit.

It would be convenient to introduce
nonlinearity parameters defined as
\begin{equation}
  \widetilde{f}_{\rm NL ({\rm new}),({\rm GR})}^{s_1 s_2 s_3}
    =30 \frac{{\widetilde{\mathcal{ A}}}_{({\rm new}),({\rm GR}) k_1=k_2=k_3}^{s_1 s_2 s_3}}{K^3},
\end{equation}
which are quantities analogous to the standard $f_{\rm NL}$
for the curvature perturbation.
We find
\begin{eqnarray}
 \widetilde{f}_{\rm NL ({\rm new})}^{s_1 s_2 s_3}
  = - \frac{5}{10368} \left[3+2(s_1 s_2 +s_2 s_3+s_3 s_1)  \right]
    \frac{H\mu}{\mathcal{ G}_T},
\end{eqnarray}
or, more concretely,
\begin{equation}
 \widetilde{f}_{\rm NL ({\rm new})}^{+++} = -\frac{5}{1152}
  \frac{H\mu}{\mathcal{ G}_T},\qquad
 \widetilde{f}_{\rm NL ({\rm new})}^{++-} = -\frac{5}{10368}
  \frac{H\mu}{\mathcal{ G}_T},
\end{equation}
with $\widetilde{f}_{\rm NL ({\rm new})}^{++-}=\widetilde{f}_{\rm NL
({\rm new})}^{+--}$ and $\widetilde{f}_{\rm NL ({\rm
new})}^{---}=\widetilde{f}_{\rm NL ({\rm new})}^{+++}$.
(This symmetry arises because parity is not violated.)
As for $\widetilde{f}_{\rm NL ({\rm GR})}^{s_1 s_2 s_3}$, we have
\begin{eqnarray}
 \widetilde{f}_{\rm NL ({\rm GR})}^{s_1 s_2 s_3}
  = \frac{85}{27648} \left[21+20(s_1 s_2 +s_2 s_3+s_3 s_1)  \right],
\end{eqnarray}
so that
\begin{equation}
 \widetilde{f}_{{\rm NL (GR})}^{+++} = \widetilde{f}_{{\rm NL (
  GR})}^{---} = \frac{255}{1024},\qquad
 \widetilde{f}_{{\rm NL (GR})}^{++-} = \widetilde{f}_{{\rm NL (
 GR})}^{+--} = \frac{85}{27648}.
\end{equation}

As defined in Eq. (\ref{eq:Bhhh}), $B^{(hhh)}_{i_1j_1i_2j_2i_3j_3}$ is
normalized by $\mathcal{ P}_h^2$. 
This normalization can be justified
when one concentrates on the non-Gaussianity of the B-mode polarization.
Because the B-mode polarization can be generated by not curvature
perturbations but tensor perturbations (except for lensing
contribution), the size of the non-Gaussianity of the B-mode
polarization could be directly characterized by $\widetilde{f}_{\rm NL
({\rm new}),({\rm GR})}^{s_1 s_2 s_3}$.

However, it should be noticed that tensor perturbations can generate not
only the B-mode polarization but also the temperature fluctuation and
the E-mode polarization. The latter two are mainly generated by the
curvature perturbations. Therefore, when one would like to quantify the
auto and cross bispectra of the temperature fluctuation and the E-mode
polarization, it would be better to normalize
$B^{(hhh)}_{i_1j_1i_2j_2i_3j_3}$ by $\mathcal{ P}_{\zeta}^2$, namely,
\begin{equation}
 B_{i_1j_1i_2j_2i_3j_3}^{(hhh)}=
  \frac{(2\pi)^4\mathcal{ P}_{\zeta}^2}{k_1^3k_2^3k_3^3}
  \left(\mathcal{ A}^{({\rm new})}_{i_1j_1i_2j_2i_3j_3}+\mathcal{ A}^{({\rm GR})}_{i_1j_1i_2j_2i_3j_3}
\right). \label{eq:Bhhh2}
\end{equation}
where $\mathcal{ A}^{({\rm new}),({\rm GR})}_{i_1j_1i_2j_2i_3j_3} = r^2
{\widetilde{\mathcal{ A}}}^{({\rm new}),({\rm GR})}_{i_1j_1i_2j_2i_3j_3}$ with
$r$ being the tensor-to-scalar ratio. In the same way, $\mathcal{ A}_{({\rm
new}),({\rm GR})}^{s_1 s_2 s_3} = r^2 {\widetilde{\mathcal{ A}}}_{({\rm new}),({\rm GR})}^{s_1 s_2
s_3}$ and $f_{\rm NL ({\rm new}),({\rm GR})}^{s_1 s_2 s_3} = r^2 \widetilde{f}_{\rm
NL ({\rm new}), ({\rm GR})}^{s_1 s_2 s_3}$.

\subsection{Two tensors and one scalar}

The cross bispectrum of two tensors and one scalar is given by
\begin{eqnarray}
\langle\zeta(\mathbf{k}_1)h_{ij}(\mathbf{k}_2)h_{kl}(\mathbf{k}_3)\rangle
=(2\pi)^3\delta(\mathbf{k}_1+\mathbf{k}_2+\mathbf{k}_3)B_{ij,kl}^{(\zeta hh)}(
\mathbf{k}_1,\mathbf{k}_2,\mathbf{k}_3),
\end{eqnarray}
where $B_{ij,kl}^{(\zeta hh)}$ is of the form:
\begin{eqnarray}
B_{ij,kl}^{(\zeta hh)}=\frac{2}{k_1^3k_2^3k_3^3}\frac{H^6}{\mathcal{ F}_S\mathcal{ F}_T^2c_sc_h^2}
\sum_{q=1}^7b_q
\mathcal{ V}_{ij,kl}^{(q)}(
\mathbf{k}_1,\mathbf{k}_2,\mathbf{k}_3)
\mathcal{ I}^{(q)}(k_1,k_2,k_3)
+(\mathbf{k}_2, i, j \leftrightarrow \mathbf{k}_3, k,l).
\end{eqnarray}
Each contribution is given by
\begin{eqnarray}
&&\mathcal{ V}_{ij,kl}^{(1)}=\Pi_{ij,mn}(\mathbf{k}_2)\Pi_{kl,mn}(\mathbf{k}_3),
\quad
\mathcal{ V}_{ij,kl}^{(2)}=
\mathbf{k}_2\cdot\mathbf{k}_3\mathcal{ V}_{ij,kl}^{(1)},
\quad
\nonumber\\&&
\mathcal{ V}_{ij,kl}^{(3)}=
\frac{\mathbf{k}_1\cdot\mathbf{k}_3}{k_1^2}\mathcal{ V}_{ij,kl}^{(1)},
\mathcal{ V}_{ij,kl}^{(4)}=
\mathcal{ V}_{ij,kl}^{(1)},
\quad
\mathcal{ V}_{ij,kl}^{(5)} = k_1^2\mathcal{ V}_{ij,kl}^{(1)},
\quad
\nonumber\\&&
\mathcal{ V}_{ij,kl}^{(6)}=\hat k_{1m}\hat k_{1n}\Pi_{ij,mm'}(\mathbf{k}_2)
\Pi_{kl,nm'}(\mathbf{k}_3),
\quad
\mathcal{ V}_{ij,kl}^{(7)} = k_1^2\mathcal{ V}_{ij,kl}^{(6)},
\end{eqnarray}
and
\begin{eqnarray}
&&\mathcal{ I}^{(1)}=\frac{1}{H^2}\frac{c_h^4k_2^2k_3^2(c_sk_1+K')}{K'^2},
\nonumber\\
&&\mathcal{ I}^{(2)}=-\frac{1}{H^2}
\frac{c_s^3k_1^3+2c_s^2c_hk_1^2(k_2+k_3)+2c_sc_h^2k_1(k_2^2+k_2k_3+k_3^2)
+c_h^3(k_2+k_3)(k_2^2+k_2k_3+k_3^2)}{K'^2},
\nonumber\\
&&\mathcal{ I}^{(3)} = \frac{1}{H^2}
\frac{c_s^2c_h^2k_1^2k_2^2(K'+c_hk_3)}{K'^2},
\quad
\mathcal{ I}^{(4)} = \frac{2}{H}\frac{c_s^2c_h^4k_1^2k_2^2k_3^2}{K'^3},
\quad
\mathcal{ I}^{(5)}=\frac{2c_h^4k_2^2k_3^2(3c_sk_1+K')}{K'^4},
\nonumber\\&&
\mathcal{ I}^{(6)} = \mathcal{ I}^{(4)},
\quad
\mathcal{ I}^{(7)} = \mathcal{ I}^{(5)},
\label{eq:shape_shh}
\end{eqnarray}
where $K':=c_sk_1+c_h(k_2+k_3)$.
Thus, it turns out that we need to evaluate only $\mathcal{ V}_{ij,kl}^{(1)}$
and $\mathcal{ V}_{ij,kl}^{(6)}$.

We would now like to define the amplitudes of the above cross bispectra
in a similar way as the case of three tensors, for which we have adopted two
different normalization conditions, (\ref{eq:Bhhh}) and (\ref{eq:Bhhh2}),
depending on whether we are interested in the B-mode polarization or the
E-mode polarization and temperature fluctuations.  The same ambiguity is
present for the cases of these cross bispectra, too.  Here we simply
normalize them in terms of $\mathcal{ P}_{\zeta}^2$ 
taking into account the fact that these bispectra generate the auto and the
cross bispectra of the temperature fluctuation and the E-mode
polarization, too,  which are mainly sourced by the curvature perturbation.
Although this normalization may not be appropriate for those including
the B-mode polarization, we do not touch the issue any further because
the change of the normalization factor from
$\mathcal{ P}_{\zeta}^2$ to $\mathcal{ P}_{\zeta}\mathcal{ P}_{h}$ or
$\mathcal{ P}_{h}^2$ can readily be done by multiplying  appropriate powers
of the tensor-to-scalar ratio $r$.  Thus we adopt the following convention:
\begin{equation}
 B_{ij,kl}^{(\zeta hh)}=
  \frac{(2\pi)^4\mathcal{ P}_{\zeta}^2}{k_1^3k_2^3k_3^3}
  \mathcal{ A}_{ij,kl}^{(\zeta hh)}, \label{eq:Bzetahh}
\end{equation}
where
\begin{eqnarray}
\mathcal{ A}_{ij,kl}^{(\zeta hh)}=8 H^2 \frac{\mathcal{ F}_S c_s}{\mathcal{ F}_T^2 c_h^2}
\sum_{q=1}^7b_q
\mathcal{ V}_{ij,kl}^{(q)}(
\mathbf{k}_1,\mathbf{k}_2,\mathbf{k}_3)
\mathcal{ I}^{(q)}(k_1,k_2,k_3)
+(\mathbf{k}_2, i, j \leftrightarrow \mathbf{k}_3, k,l).
\end{eqnarray}
We also define the following cross bispectra:
\begin{equation}
  \langle\zeta(\mathbf{k}_1)\xi^{(s_2)}(\mathbf{k}_2)\xi^{(s_3)}(\mathbf{k}_3)\rangle
=(2\pi)^3\delta(\mathbf{k}_1+\mathbf{k}_2+\mathbf{k}_3)B_{s2,s3}^{(\zeta hh)}(
\mathbf{k}_1,\mathbf{k}_2,\mathbf{k}_3).
\end{equation}
Here $B_{s_2,s_3}^{(\zeta hh)}$ and $\mathcal{ A}_{s_2,s_3}^{(\zeta hh)}$ are given by
\begin{eqnarray}
B_{s2,s3}^{(\zeta hh)}&=&\frac{(2\pi)^4\mathcal{ P}_{\zeta}^2}{k_1^3k_2^3k_3^3}
  \mathcal{ A}^{(\zeta hh)}_{s_2,s_3} \nonumber\\
&=&
\frac{2}{k_1^3k_2^3k_3^3}\frac{H^6}{\mathcal{ F}_S\mathcal{ F}_T^2c_sc_h^2}
\sum_{q=1}^7b_q
\mathcal{ V}_{s2,s3}^{(q)}(
\mathbf{k}_1,\mathbf{k}_2,\mathbf{k}_3)
\mathcal{ I}^{(q)}(k_1,k_2,k_3)
+(\mathbf{k}_2, s_2 \leftrightarrow \mathbf{k}_3, s_3), \nonumber\\
\end{eqnarray}
where $\mathcal{ V}_{s2,s3}^{(q)}( \mathbf{k}_1,\mathbf{k}_2,\mathbf{k}_3)$ is
evaluated as
\begin{eqnarray}
&&\mathcal{ V}_{s2,s3}^{(1)} =  \frac{1}{16 k_2^2 k_3^2}
         \left[k_1^2-(s_2k_2+ s_3 k_3)^2 \right]^2,
		 \quad
\mathcal{ V}_{s2,s3}^{(2)} =  \mathbf{k}_2\cdot\mathbf{k}_3\mathcal{ V}_{s2,s3}^{(1)}
               = \frac{k_1^2-k_2^2-k_3^2}{2} \mathcal{ V}_{s2,s3}^{(1)},
			   \nonumber\\&&
\mathcal{ V}_{s2,s3}^{(3)} = \frac{\mathbf{k}_1\cdot\mathbf{k}_3}{k_1^2}
                    \mathcal{ V}_{s2,s3}^{(1)}
               = -\frac{k_1^2-k_2^2+k_3^2}{2 k_1^2} \mathcal{ V}_{s2,s3}^{(1)},
			   \quad
\mathcal{ V}_{s2,s3}^{(4)} =  \mathcal{ V}_{s2,s3}^{(1)},
\quad
\mathcal{ V}_{s2,s3}^{(5)} =  k_1^2\mathcal{ V}_{s2,s3}^{(1)},
\nonumber\\&&
\mathcal{ V}_{s2,s3}^{(6)} =  \frac{K}{32 k_1^2 k_2^2 k_3^2}
    (k_1-k_2-k_3)(k_1+k_2-k_3)(k_1-k_2+k_3)
    \left[k_1^2-(s_2 k_2+s_3 k_3)^2 \right],\quad
    \nonumber\\&&
\mathcal{ V}_{s2,s3}^{(7)}= k_1^2\mathcal{ V}_{s2,s3}^{(6)}.
\end{eqnarray}

\subsection{Two scalars and one tensor}

The cross bispectrum of two scalars and one tensor is given by
\begin{eqnarray}
\langle\zeta(\mathbf{k}_1)\zeta(\mathbf{k}_2)h_{ij}(\mathbf{k}_3)\rangle
=(2\pi)^3\delta(\mathbf{k}_1+\mathbf{k}_2+\mathbf{k}_3)B_{ij}^{(\zeta \zeta h)}(
\mathbf{k}_1,\mathbf{k}_2,\mathbf{k}_3),
\end{eqnarray}
where $B_{ij}^{(\zeta \zeta h)}$ is of the form
\begin{eqnarray}
B_{ij}^{(\zeta \zeta h)}=\frac{1}{4k_1^3k_2^3k_3^3}\frac{H^6}{\mathcal{ F}_S^2\mathcal{ F}_Tc_s^2c_h}
\sum_{q=1}^6c_q
\mathcal{ V}_{ij}^{(q)}(
\mathbf{k}_1,\mathbf{k}_2,\mathbf{k}_3)
\mathcal{ J}^{(q)}(k_1,k_2,k_3)
+(\mathbf{k}_1 \leftrightarrow \mathbf{k}_2).
\end{eqnarray}
Each contribution is given by
\begin{eqnarray}
&&\mathcal{ V}_{ij}^{(1)} = k_{1k}k_{2l}\Pi_{ij,kl}(\mathbf{k}_3),
\quad
\mathcal{ V}_{ij}^{(2)}=\mathcal{ V}_{ij}^{(1)},
\quad
\mathcal{ V}_{ij}^{(3)} = \frac{1}{k_2^2}\mathcal{ V}_{ij}^{(1)},
\quad
\nonumber\\&&
\mathcal{ V}_{ij}^{(4)}=\frac{k_3^2}{k_2^2}\mathcal{ V}_{ij}^{(1)},
\quad
\mathcal{ V}_{ij}^{(5)} =k_3^2\mathcal{ V}_{ij}^{(1)},
\quad
\mathcal{ V}_{ij}^{(6)}=\frac{k_3^2}{k_1^2k_2^2}\mathcal{ V}_{ij}^{(1)},
\end{eqnarray}
and
\begin{eqnarray}
&&\mathcal{ J}^{(1)}=-\frac{1}{H^2}
\frac{c_s^3(k_1+k_2)(k_1^2+k_1k_2+k_2^2)+2c_s^2c_h(k_1^2+k_1k_2+k_2^2)k_3
+2c_sc_h^2(k_1+k_2)k_3^2+c_h^3k_3^3}
{K''^2},
\nonumber\\
&&\mathcal{ J}^{(2)}=\frac{1}{H}\frac{c_h^2k_3^2
[2c_s^2(k_1^2+3k_1k_2+k_2^2)+3c_sc_h(k_1+k_2)k_3+c_h^2k_3^2]
}{K''^3},
\nonumber\\
&&\mathcal{ J}^{(3)}=\frac{1}{H^2}\frac{c_s^2c_h^2k_2^2k_3^2(c_sk_1+K'')}{K''^2},
\nonumber\\
&&
\mathcal{ J}^{(4)}=\frac{1}{H}\frac{c_s^2k_2^2
[c_s^2(k_1+k_2)(2k_1+k_2)+3c_sc_h(2k_1+k_2)k_3+2c_h^2k_3^2]}
{K''^3},
\nonumber\\
&&
\mathcal{ J}^{(5)}=\frac{2}{K''^4}
\left[
c_s^3(k_1+k_2)(k_1^2+3k_1k_2+k_2^2)+4c_s^2c_h(k_1^2+3k_1k_2+k_2^2)k_3
+4c_sc_h^2(k_1+k_2)k_3^2+c_h^3k_3^3
\right],
\nonumber\\
&&
\mathcal{ J}^{(6)}=\frac{1}{H^2}\frac{c_s^4k_1^2k_2^2(K''+c_hk_3)}{K''^2},
\label{eq:shape_ssh}
\end{eqnarray}
with $K'':=c_s(k_1+k_2)+c_hk_3$.
Thus, it turns out that we need to evaluate only $\mathcal{ V}_{ij}^{(1)}$.

As in the case of two tensors and one scalar, we normalize
the bispectrum by $\mathcal{ P}_{\zeta}^2$ as
\begin{equation}
 B_{ij}^{(\zeta \zeta h)}=
  \frac{(2\pi)^4\mathcal{ P}_{\zeta}^2}{k_1^3k_2^3k_3^3}
  \mathcal{ A}_{ij}^{(\zeta \zeta h)}, \label{eq:Bzetazetah}
\end{equation}
where
\begin{eqnarray}
\mathcal{ A}_{ij}^{(\zeta \zeta h)}=\frac{H^2}{\mathcal{ F}_T c_h}
\sum_{q=1}^6c_q
\mathcal{ V}_{ij}^{(q)}(
\mathbf{k}_1,\mathbf{k}_2,\mathbf{k}_3)
\mathcal{ J}^{(q)}(k_1,k_2,k_3)
+(\mathbf{k}_1 \leftrightarrow \mathbf{k}_2),
\end{eqnarray}

We also define the following cross bispectra:
\begin{equation}
  \langle\zeta(\mathbf{k}_1)\zeta(\mathbf{k}_2)\xi^{(s)}(\mathbf{k}_3)\rangle
=(2\pi)^3\delta(\mathbf{k}_1+\mathbf{k}_2+\mathbf{k}_3)B_{s}^{(\zeta
\zeta h)}(\mathbf{k}_1,\mathbf{k}_2,\mathbf{k}_3).
\end{equation}
Here $B_{s}^{(\zeta \zeta h)}$ and $A_{s}^{(\zeta \zeta h)}$ are given by
\begin{eqnarray}
B_{s}^{(\zeta \zeta h)}&=&\frac{(2\pi)^4\mathcal{ P}_{\zeta}^2}{k_1^3k_2^3k_3^3}
  \mathcal{ A}^{(\zeta \zeta h)}_{s} \nonumber\\
&=& \frac{1}{4k_1^3k_2^3k_3^3}\frac{H^6}{\mathcal{ F}_S^2\mathcal{ F}_Tc_s^2c_h}
\sum_{q=1}^6c_q
\mathcal{ V}_{s}^{(q)}(
\mathbf{k}_1,\mathbf{k}_2,\mathbf{k}_3)
\mathcal{ J}^{(q)}(k_1,k_2,k_3)
+(\mathbf{k}_1 \leftrightarrow \mathbf{k}_2),
\end{eqnarray}
where $\mathcal{ V}_{s}^{(q)}( \mathbf{k}_1,\mathbf{k}_2,\mathbf{k}_3)$ is
evaluated as
\begin{eqnarray}
\mathcal{ V}_{s}^{(1)}= \frac{K}{8k_3^2}
         (k_1-k_2-k_3)(k_1+k_2-k_3)(k_1-k_2+k_3),
\end{eqnarray}
and
\begin{eqnarray}
\mathcal{ V}_{s}^{(2)}= \mathcal{ V}_{s}^{(1)},
\quad
\mathcal{ V}_{s}^{(3)}= \frac{1}{k_2^2} \mathcal{ V}_{s}^{(1)},
\quad
\mathcal{ V}_{s}^{(4)}= \frac{k_3^2}{k_2^2} \mathcal{ V}_{s}^{(1)},
\quad
\mathcal{ V}_{s}^{(5)}= k_3^2\mathcal{ V}_{s}^{(1)},
\quad
\mathcal{ V}_{s}^{(6)}= \frac{k_3^2}{k_1^2 k_2^2} \mathcal{ V}_{s}^{(1)}.
\end{eqnarray}
Indeed, the above functions are independent of $s$ due to no parity
violation.

\subsection{Three scalars}

Here we give the bispectrum defined by
\begin{eqnarray}
\langle\zeta(\mathbf{k}_1)\zeta(\mathbf{k}_2)\zeta(\mathbf{k}_3)\rangle
=(2\pi)^3\delta(\mathbf{k}_1+\mathbf{k}_2+\mathbf{k}_3)B^{(\zeta\zeta\zeta)}(k_1,k_2,k_3).
\end{eqnarray}
The result is given in Ref.~\cite{Gao:2011qe, DeFelice:2011uc}: 
\begin{eqnarray}
B^{(\zeta\zeta\zeta)}&=&
\frac{\left(2\pi\right)^{4}\mathcal{ P}_{\zeta}^{2}}
{4k_{1}^{3}k_{2}^{3}k_{3}^{3}}
\left[
\frac{(k_{1}k_{2}k_{3})^{2}}{K^{3}}\mathcal{C}_{1}
+\frac{\mathcal{C}_{2}}{K}\left(2\sum_{i>j}k_{i}^{2}k_{j}^{2}-\frac{1}{K}
\sum_{i\neq j}k_{i}^{2}k_j^3\right)\right.
\nonumber \\ &&\quad
+\mathcal{C}_{3}\left(\sum_{i}k_{i}^{3}+\frac{4}{K}\sum_{i>j}
k_{i}^{2}k_{j}^{2}-\frac{2}{K^{2}}\sum_{i\neq j}k_{i}^{2}k_{j}^{3}\right) \nonumber\\&&
+\mathcal{C}_{4}\left(\sum_{i}k_{i}^{3}-\frac{1}{2}\sum_{i\neq j}k_{i}k_{j}^{2}
-\frac{2}{K^{2}}\sum_{i\neq j}k_{i}^{2}k_{j}^{3}\right)
\nonumber \\&&\quad
\left.
+\frac{\mathcal{C}_{5}}{K^{2}}\left(2\sum_{i}k_{i}^{5}
+\sum_{i\neq j}k_{i}k_{j}^{4}-3\sum_{i\neq j}k_{i}^{2}k_{j}^{3}-2k_{1}k_{2}k_{3}
\sum_{i>j}k_{i}k_{j}\right)\right].\label{zeta_bi}
\end{eqnarray}

\subsection{
Shapes of the cross bispectra in momentum space
}

\begin{figure}[htbp]
\begin{center}
\subfigure[]{   \includegraphics[width=55mm]{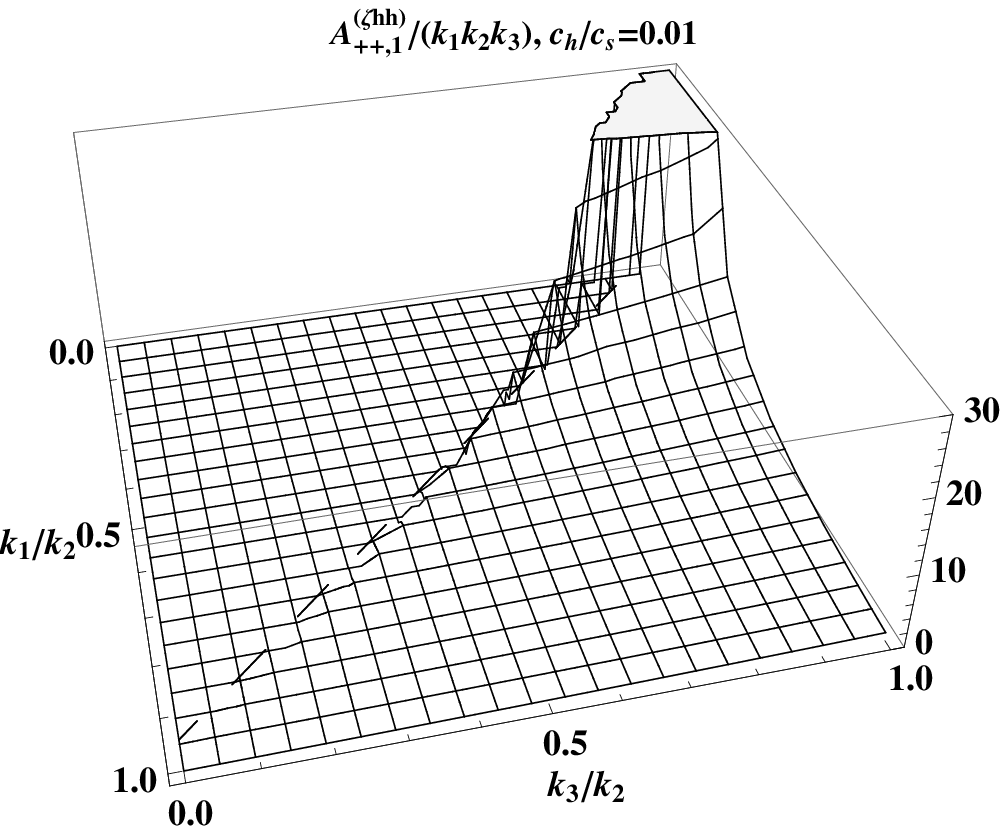}}
\subfigure[]{ \includegraphics[width=55mm]{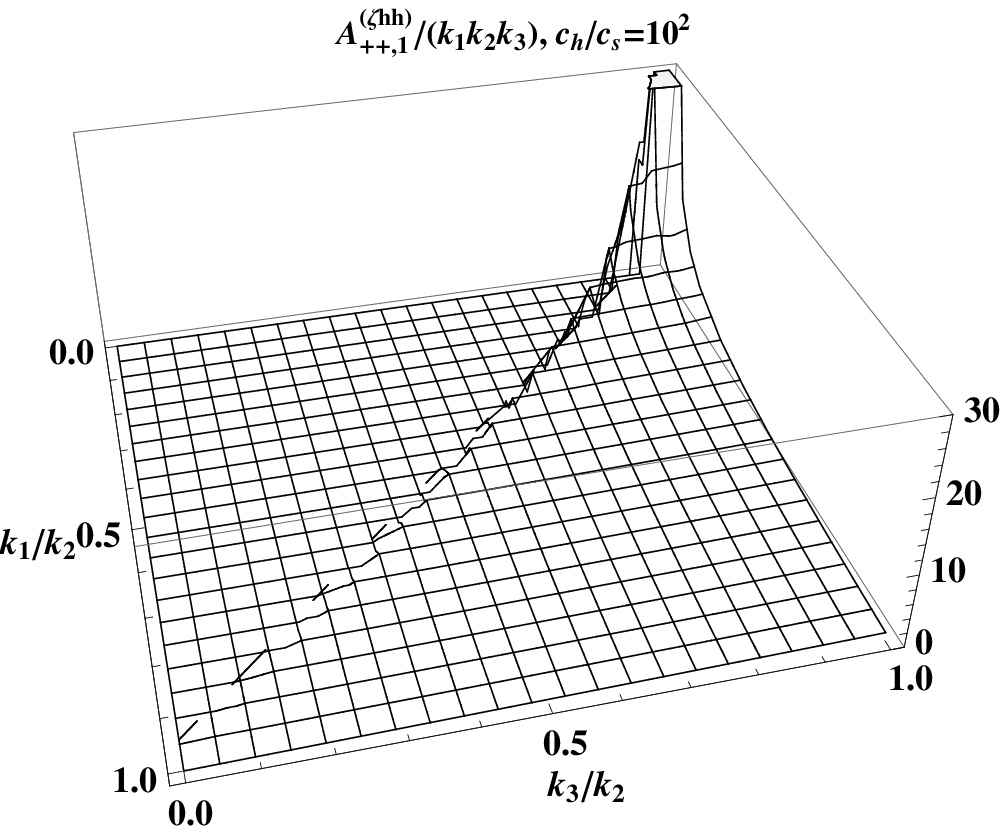}}
\end{center}
\caption{
$\mathcal{ A}_{++,1}^{(\zeta h h )}(k_1k_2k_3)^{-1}$
as a function of $k_1 / k_2$ and $k_3 / k_2$
for $c_h / c_s = 0.01$ in (a) and $c_h / c_s = 10^2$ in (b), normalized to unity for $k_1 = k_2 = k_3$.}
\label{fig:STT_bis_1.eps}
\end{figure}
\begin{figure}[htbp]
\begin{center}
\subfigure[]{   \includegraphics[width=55mm]{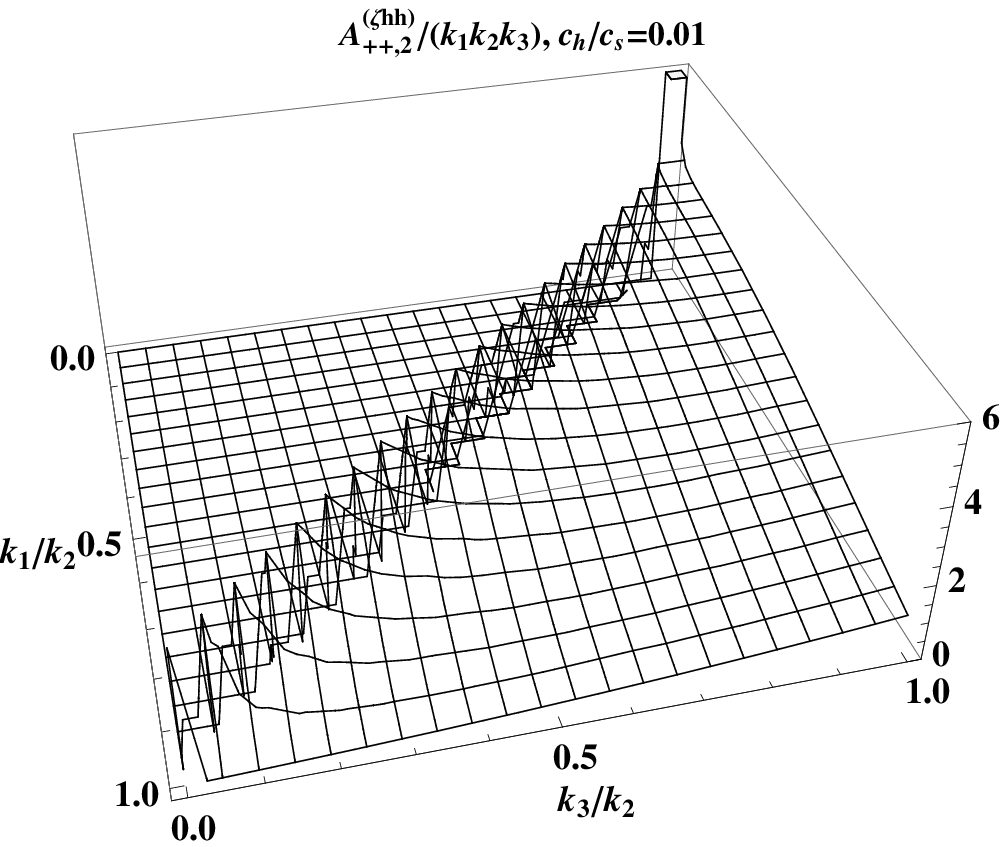}}
\subfigure[]{ \includegraphics[width=55mm]{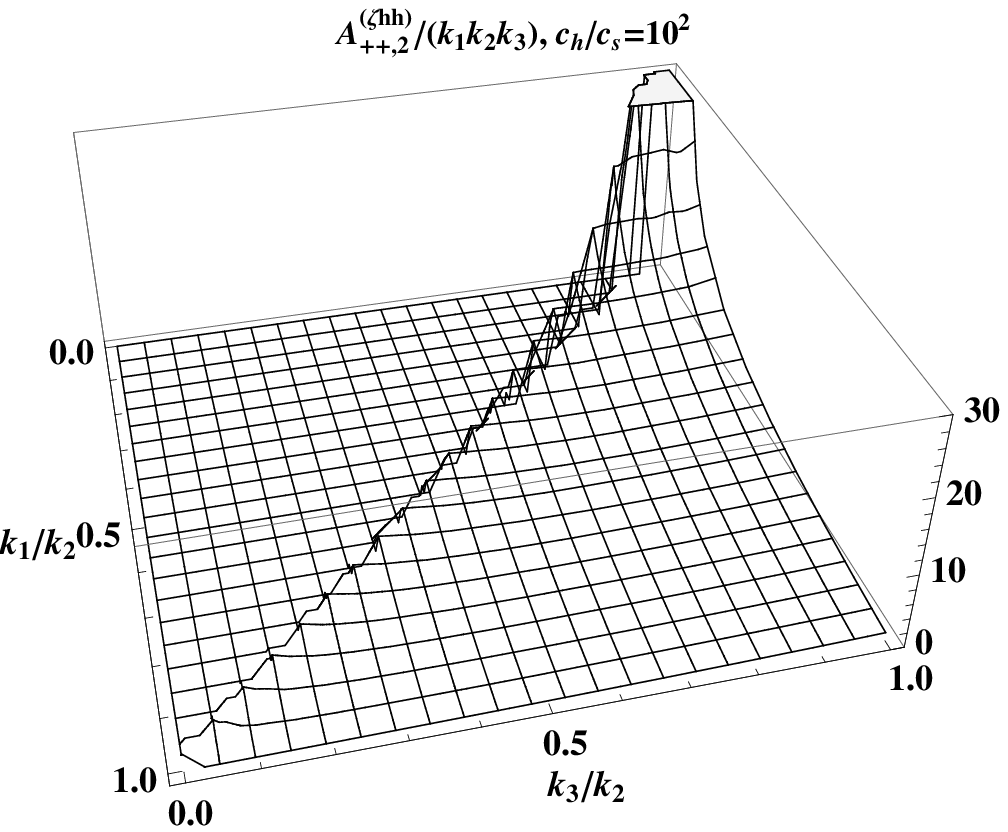}}
\end{center}
\caption{
$\mathcal{ A}_{++,2}^{(\zeta h h )}(k_1k_2k_3)^{-1}$
as a function of $k_1 / k_2$ and $k_3 / k_2$
for $c_h / c_s = 0.01$ in (a) and $c_h / c_s = 10^2$ in (b), normalized to unity for $k_1 = k_2 = k_3$.}
\label{fig:STT_bis_2.eps}
\end{figure}
\begin{figure}[htbp]
\begin{center}
\subfigure[]{   \includegraphics[width=55mm]{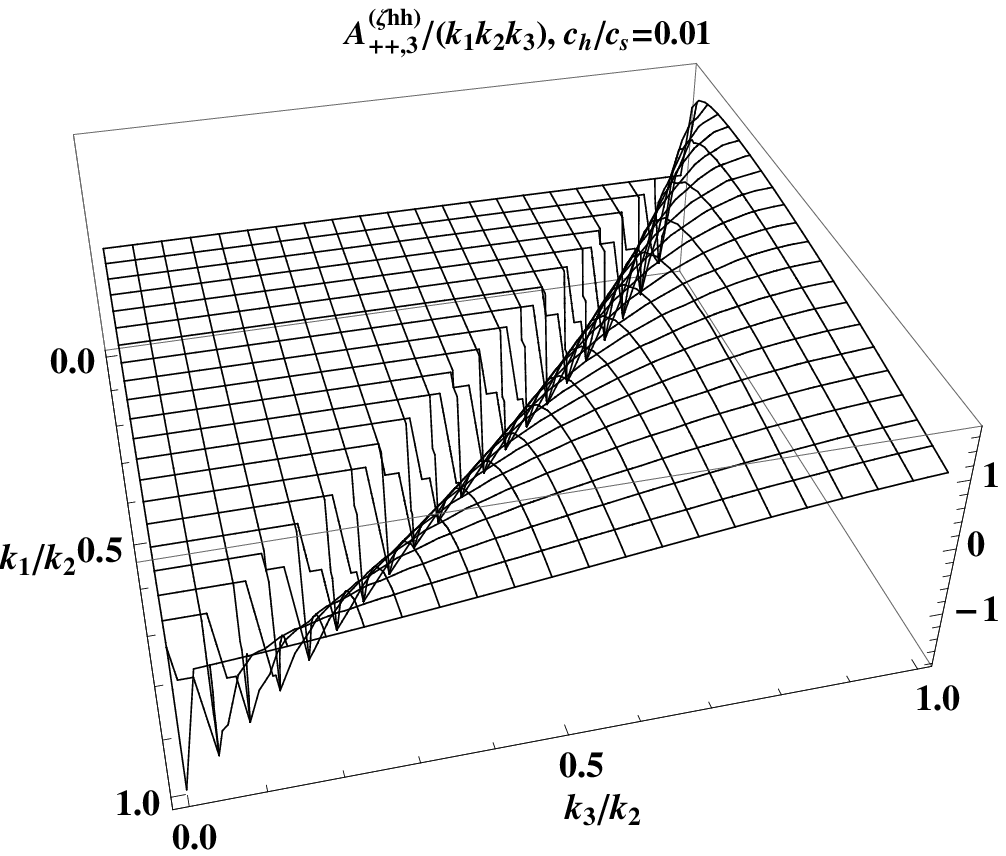}}
\subfigure[]{ \includegraphics[width=55mm]{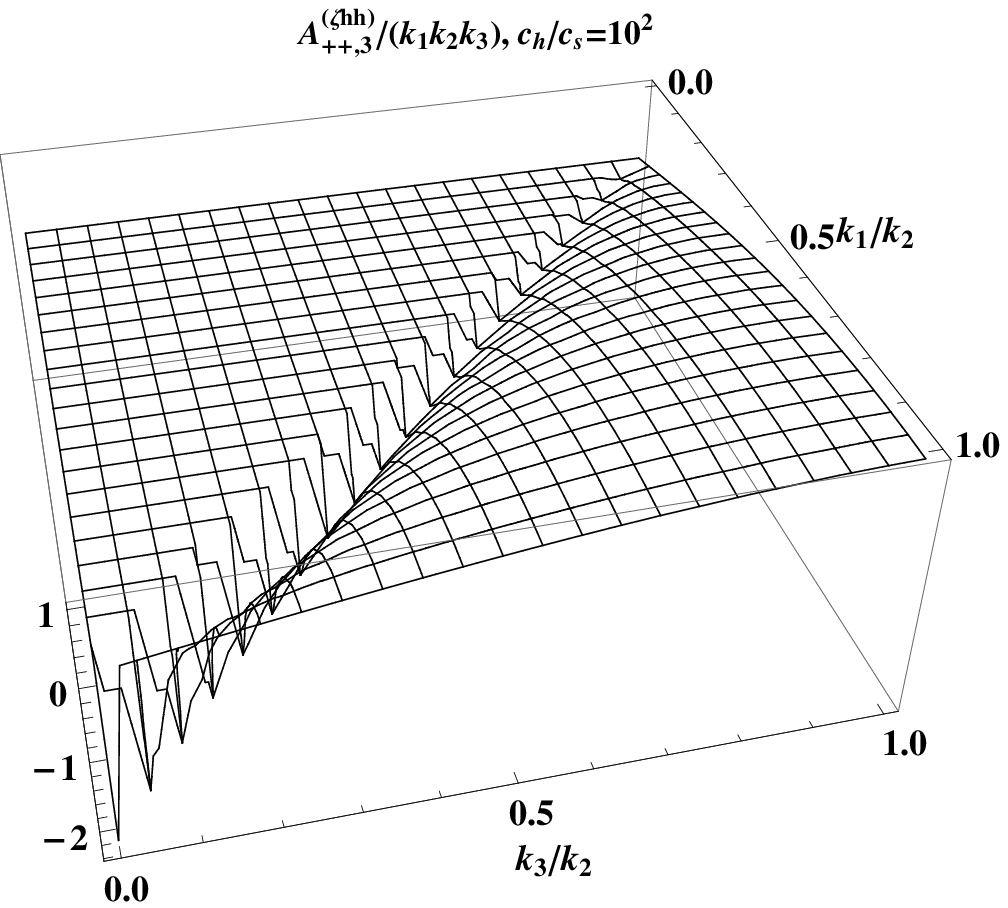}}
\end{center}
\caption{
$\mathcal{ A}_{++,3}^{(\zeta h h )}(k_1k_2k_3)^{-1}$
as a function of $k_1 / k_2$ and $k_3 / k_2$
for $c_h / c_s = 0.01$ in (a) and $c_h / c_s = 10^2$ in (b), normalized to unity for $k_1 = k_2 = k_3$.}
\label{fig:STT_bis_3.eps}
\end{figure}

\begin{figure}[htbp]
\begin{center}
\subfigure[]{   \includegraphics[width=55mm]{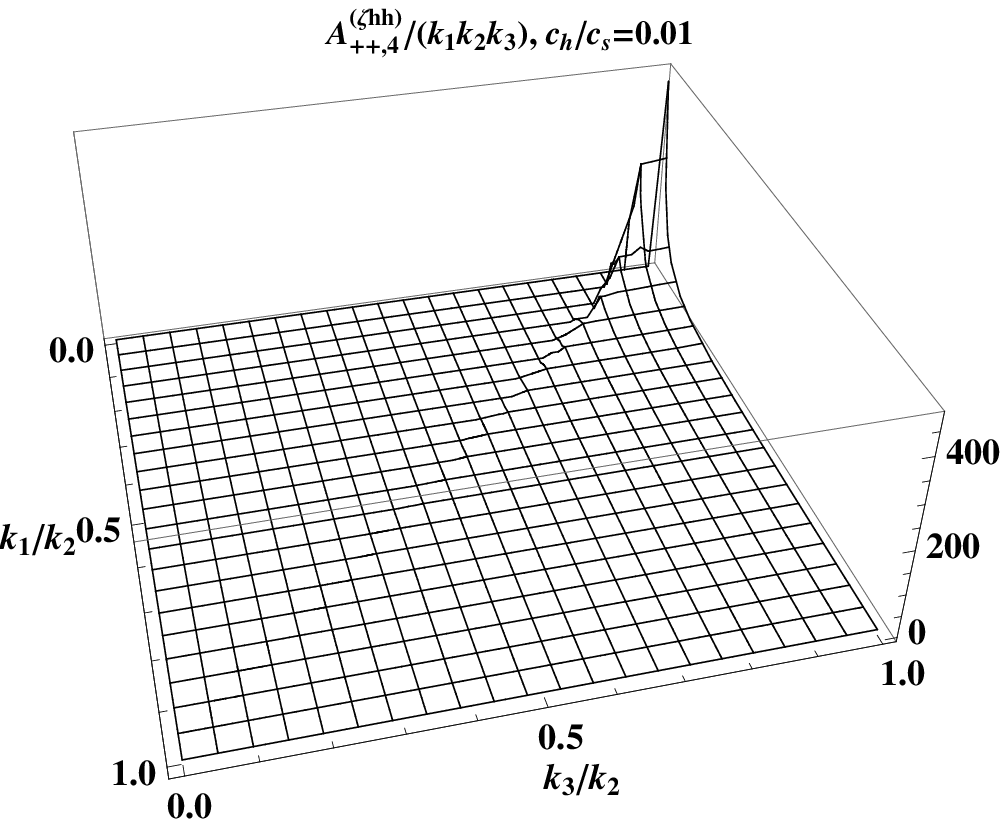}}
\subfigure[]{ \includegraphics[width=55mm]{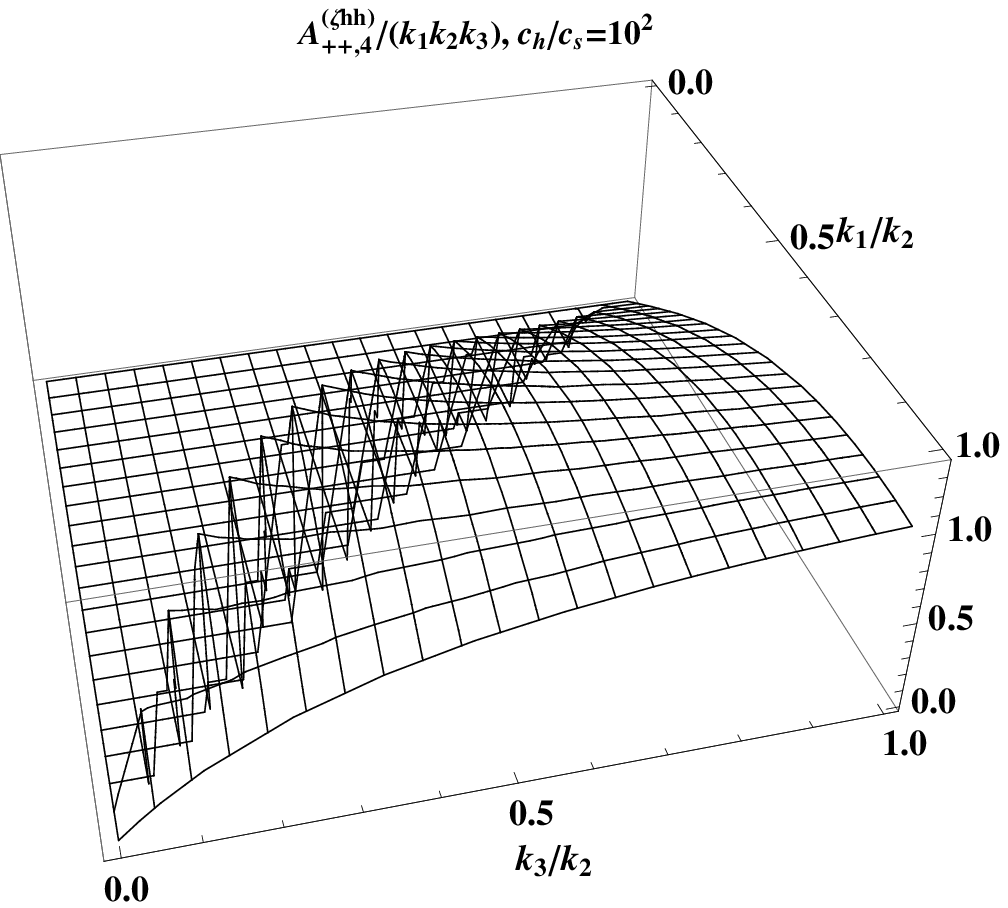}}
\end{center}
\caption{
$\mathcal{ A}_{++,4}^{(\zeta h h )}(k_1k_2k_3)^{-1}$
as a function of $k_1 / k_2$ and $k_3 / k_2$
for $c_h / c_s = 0.01$ in (a) and $c_h / c_s = 10^2$ in (b), normalized to unity for $k_1 = k_2 = k_3$.}
\label{fig:STT_bis_4.eps}
\end{figure}

Let us discuss the shape of each cross bispectrum in momentum space.
As shown in Ref. \cite{gw-non-g} and also mentioned in the previous subsection,
for the bispectrum of the tensor mode,
$ {\widetilde{\mathcal{ A}}}_{({\rm new})}^{+++}$ and $ {\widetilde{\mathcal{ A}}}_{({\rm GR})}^{+++}$
have respectively peaks in the equilateral and squeezed limits.
The shape of the bispectrum of scalar perturbations was also discussed in
Ref. \cite{Gao:2011qe, DeFelice:2011uc} and the authors have found that it is well approximated
by the equilateral shape. 

In a similar way to the auto bispectra of tensors and scalars,
we can also discuss the shapes
of the cross bispectra of tensors and scalars
 in momentum space.
However, contrary to the auto-bispectra of tensors and scalars,
the shapes of cross bispectra strongly depend on the sound speeds of the tensor and scalar perturbations,
as can be seen in Eqs. (\ref{eq:shape_shh}) and (\ref{eq:shape_ssh}). Here, we denote a term proportional to $b_q$ in $\mathcal{ A}_{s_2,s_3}^{(\zeta h h)}$ as
$\mathcal{ A}_{s_2,s_3,q}^{(\zeta h h)}$
and also a term proportional to $c_q$ in $\mathcal{ A}_{s}^{(\zeta \zeta h)}$ as
$\mathcal{ A}_{s,q}^{(\zeta \zeta h)}$.
The shape of $\mathcal{ A}_{++,1}^{(\zeta h h )}(k_1k_2k_3)^{-1}$ in $k$-space for two limiting cases
is plotted in Fig. \ref{fig:STT_bis_1.eps}. The left panel (a) and the right one (b) are respectively
for the cases with $c_h / c_s = 0.01$ and $c_h / c_s = 10^2$. This figure implies that $\mathcal{ A}_{++,1}^{(\zeta h h )}(k_1k_2k_3)^{-1}$ has a peak in the squeezed limit ($k_1 \ll k_2 \sim k_3$) for both limiting cases. However, the sharpness of the peak seems to depend on the value of $c_h / c_s$.
In Fig. \ref{fig:STT_bis_2.eps} where $\mathcal{ A}_{++,2}^{(\zeta h h )}(k_1k_2k_3)^{-1}$
is plotted, we find that $\mathcal{ A}_{++,2}^{(\zeta h h )}(k_1k_2k_3)^{-1}$
for the case with $c_h / c_s = 0.01$
has a sharp peak in the squeezed limit together 
with a non-trivial shape in wide region of the momentum space. For the case with $c_h / c_s = 10^2$ (shown in Fig. \ref{fig:STT_bis_2.eps}-(b)), $\mathcal{ A}_{++,2}^{(\zeta h h )}(k_1k_2k_3)^{-1}$ also has a peak in the squeezed limit.

 Contrary to $\mathcal{ A}_{++,1}^{(\zeta h h )}(k_1k_2k_3)^{-1}$
and  $\mathcal{ A}_{++,2}^{(\zeta h h )}(k_1k_2k_3)^{-1}$, both of which have a peak in the squeezed limit,
$\mathcal{ A}_{++,3}^{(\zeta h h )}(k_1k_2k_3)^{-1}$ does not have any sharp peak, but
its shape strongly depends on the value of $c_h / c_s$, as shown in Fig. \ref{fig:STT_bis_3.eps}.
In the case with $c_h / c_s \ll 1$, $\mathcal{ A}_{++,3}^{(\zeta h h )}(k_1k_2k_3)^{-1}$
becomes large at $k_1 \ll k_2$, and then
its shape looks to come close to so-called orthogonal type in the limit of $c_h / c_s \gg 1$.
$\mathcal{ A}_{++,4}^{(\zeta h h )}(k_1k_2k_3)^{-1}$ also strongly depends on the value of $c_h/c_s$.
As can be seen in Fig. \ref{fig:STT_bis_4.eps}, the peak of $\mathcal{ A}_{++,4}^{(\zeta h h )}(k_1k_2k_3)^{-1}$ 
shifts in the momentum space depending on $c_h / c_s$, and 
$\mathcal{ A}_{++,4}^{(\zeta h h )}(k_1k_2k_3)^{-1}$ for small $c_h$ has a finite value even in the squeezed limit. Although we do not show here,
we also found that $\mathcal{ A}_{++,5}^{(\zeta h h )}$, $\mathcal{ A}_{++,6}^{(\zeta h h )}$
and $\mathcal{ A}_{++,7}^{(\zeta h h )}$
 have almost same shapes as
$\mathcal{ A}_{++,4}^{(\zeta h h )}$.

\begin{figure}[htbp]
\begin{center}
\subfigure[]{   \includegraphics[width=55mm]{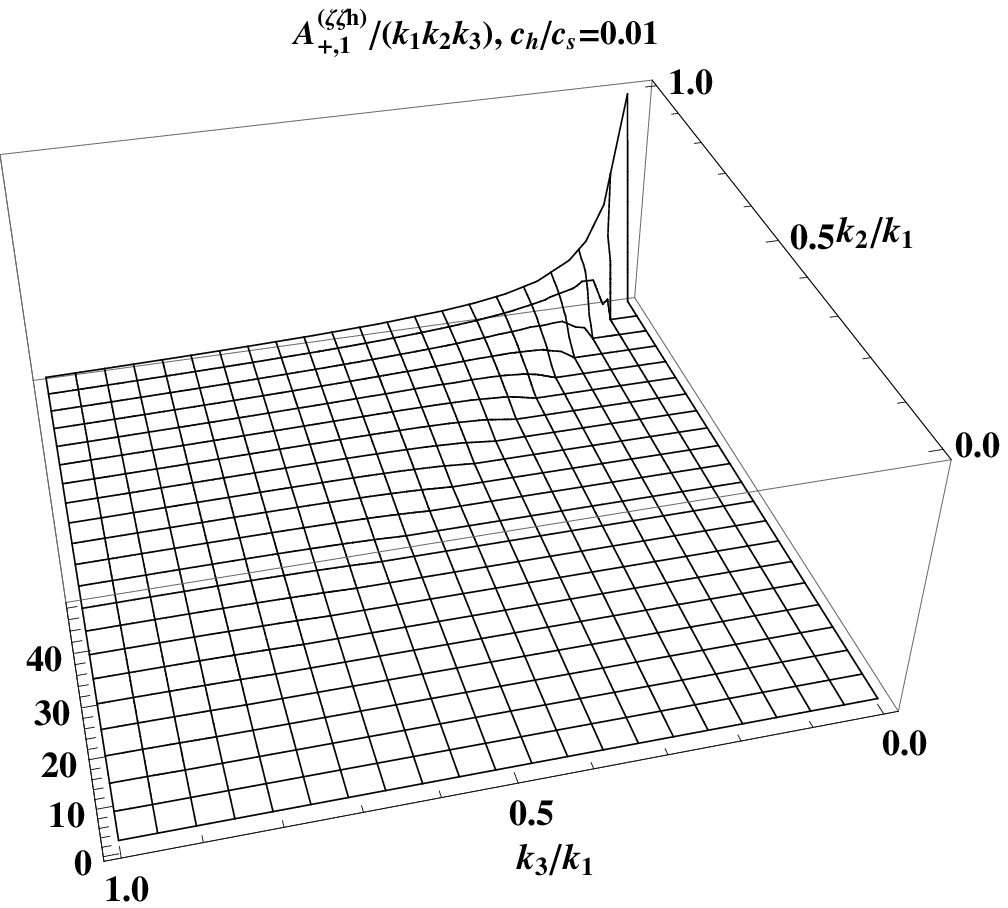}}
\subfigure[]{ \includegraphics[width=55mm]{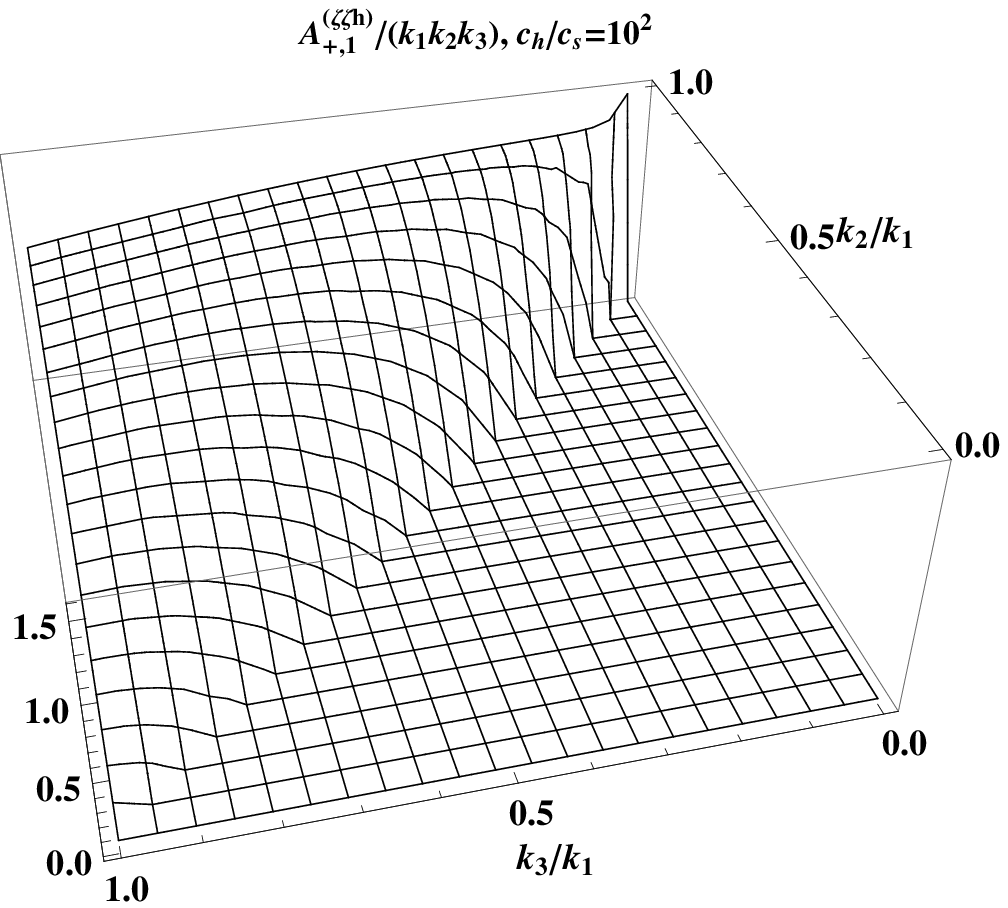}}
\end{center}
\begin{center}
\subfigure[]{   \includegraphics[width=55mm]{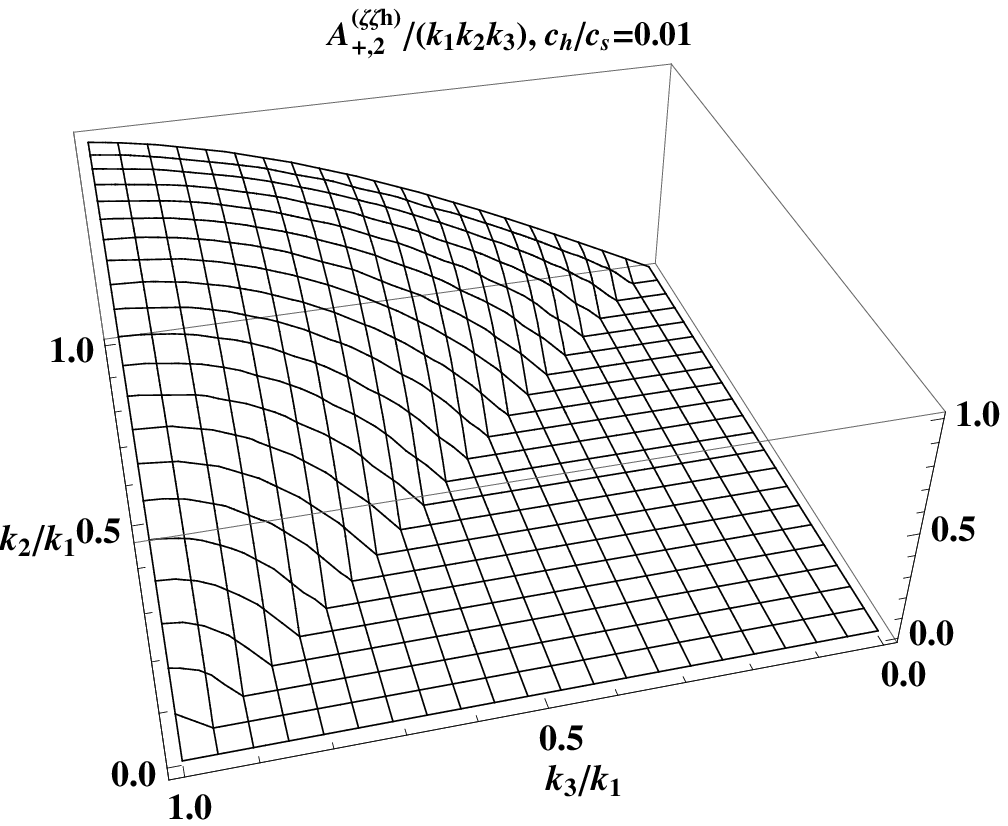}}
\subfigure[]{ \includegraphics[width=55mm]{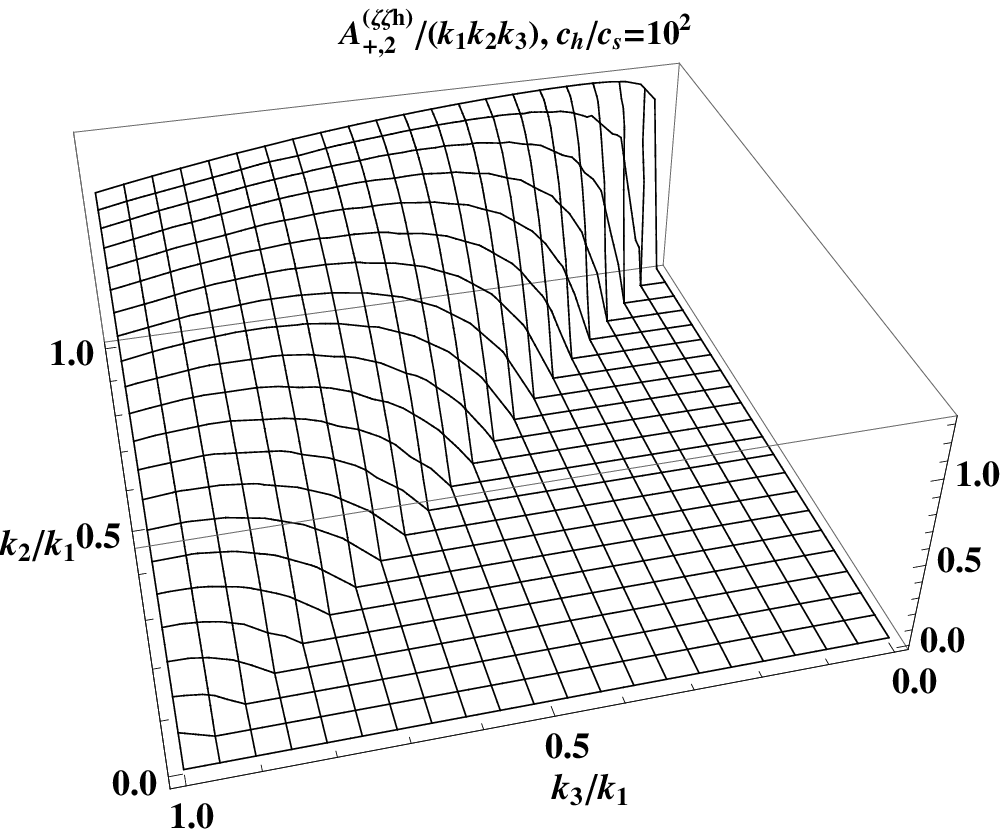}}
\end{center}
\caption{
$\mathcal{ A}_{+,q}^{(\zeta \zeta h )}(k_1k_2k_3)^{-1}$ ((a) and (b) for $q=1$, and (c) and (d) for $q=2$)
as a function of $k_2 / k_1$ and $k_3 / k_1$
for $c_h / c_s = 0.01$ in (a) and (c), and $c_h / c_s = 10^2$ in (b) and (d), normalized to unity for $k_1 = k_2 = k_3$.}
\label{fig:SST_bis_1.eps}
\end{figure}

In Fig. \ref{fig:SST_bis_1.eps}, $\mathcal{ A}_{+,q}^{(\zeta \zeta h )}(k_1k_2k_3)^{-1}$ for $q=1,2$ are  plotted.  
This figure shows that $\mathcal{ A}_{+,q}^{(\zeta \zeta h )}(k_1k_2k_3)^{-1}$ also has strong dependence
on $c_h / c_s$ and there is no divergence feature in the whole region of the momentum space, unlike the
so-called local shape. 
Since we found $\mathcal{ A}_{+,q}^{(\zeta \zeta h )}$ for $q=3,4,5,6$
 have almost same shapes as
$\mathcal{ A}_{+,2}^{(\zeta \zeta h )}$, we do not show the plots for these contributions.

The detailed analysis of the shapes of the cross bispectra, including a precise comparison
with the standard local-, equilateral and orthogonal shapes,
is an issue in progress with the detailed analysis of CMB bispectra \cite{CMBbispectra}.


\section{Examples}

In this section, we consider two representative examples of inflation to
estimate the amount of non-Gaussianities from tensor and scalar
perturbations.  The first example is general potential-driven inflation
studied in Ref.~\cite{GHiggs}.  This class of inflation models includes
variants of Higgs inflation enabled by enhancing the effect of Hubble
friction. These potential driven models have $c_s^2=\mathcal{ O}(1)$ and
$c_h^2\simeq 1$.  Next, to see the impact of generic $c_s^2$ more clearly, we
study k-inflation as another  example.

\subsection{The case of potential-driven inflation models}

We wish to treat a wide class of potential-driven inflation models at
one time.  For this purpose, we introduce six $\phi$-dependent functions
to write
\begin{eqnarray}
K=-V(\phi)+\mathcal{ K}(\phi)X,
\quad
G_3=h_3(\phi)X,
\quad
G_4=g(\phi)+h_4(\phi)X,
\quad
G_5=h_5(\phi)X.\label{ghiggsfns}
\end{eqnarray}
In particular, the above form includes different Higgs inflation models
proposed so far~\cite{GHiggs}.
These may also be regarded as the Taylor expansion of
$K(\phi, X)$ and $G_i(\phi, X)$ with respect to $X$.
Would-be leading terms in $G_3$ and $G_5$ have been 
removed without loss of generality.

Slow-roll dynamics of general potential-driven inflation models
has been addressed in Ref.~\cite{GHiggs}. During inflation
we assume that the following slow-roll conditions are satisfied:
\begin{eqnarray}
&&\epsilon=-\frac{\dot H}{H^2}\ll 1,\quad
\eta=-\frac{\ddot\phi}{H\dot\phi}\ll 1,\quad
\delta=\frac{\dot g}{Hg}\ll 1,\quad \nonumber\\&&
\alpha_2=\frac{\dot{\mathcal{ K}}}{H\mathcal{ K}}\ll 1,
\quad
\alpha_i=\frac{\dot h_i}{Hh_i}\ll 1\;\;(i=3,4,5).
\end{eqnarray}
It is convenient to define
\begin{eqnarray}
u(\phi):=\mathcal{ K}+\frac{h_4V}{g},
\quad
v(\phi):=h_3+\frac{h_5V}{6g},
\quad
W(\phi):=\frac{1}{2}\left[u+
\sqrt{u^2-4g^2v\frac{\D}{\D\phi}\left(\frac{V}{g^2}\right)}\right].
\end{eqnarray}
Under the slow-roll approximation the gravitational field equations reduce to
\begin{eqnarray}
6gH^2\simeq V,\quad
2\epsilon+\delta\simeq
\frac{X}{gH^2}\left( u+3H  \dot\phi v\right).
\end{eqnarray}

Now it is easy to see that $\mathcal{ F}_T\simeq \mathcal{ G}_T\simeq 2g$ and
\begin{eqnarray}
&&\mathcal{ F}_S\simeq\frac{X}{H^2}\left(u+4H\dot\phi v\right)
\simeq\frac{g}{3}(2\epsilon +\delta)\left(4-\frac{u}{W}\right),
\quad
\nonumber\\&&
\mathcal{ G}_S\simeq\frac{X}{H^2}\left(u+6H\dot\phi v\right)
\simeq g(2\epsilon +\delta)\left(2-\frac{u}{W}\right),
\end{eqnarray}
so that $\mathcal{ F}_S$ and $\mathcal{ G}_S$ are slow-roll suppressed.
It can also be seen that $c_h^2\simeq 1$ and $c_s^2=\mathcal{ O}(1)$.

The coefficients of the cubic terms are given by
\begin{eqnarray}
&&
b_1\simeq\frac{Xu}{8H^2},
\quad
b_2\simeq \frac{g}{24}(2\epsilon+\delta)\left(4-\frac{u}{W}\right),
\quad
b_3\simeq -\frac{g}{4}(2\epsilon+\delta)\left(2-\frac{u}{W}\right),
\nonumber\\&&
b_4\simeq -\mu,
\quad
b_5\simeq -\frac{\mu}{6H}\frac{1-u/W}{2-u/W},
\quad
 b_7\simeq \frac{\mu}{2H},
 \quad
 E_{shh}\simeq\frac{1}{4H}\zeta\dot h_{ij}E_{ij}^h,
\end{eqnarray}
and
\begin{eqnarray}
&&c_1\simeq \frac{g}{3}(2\epsilon+\delta)\left(4-\frac{u}{W}\right),
\quad
c_2\simeq \frac{g}{12H}(2\epsilon+\delta)\left(1-\frac{u}{W}\right)
+\frac{\dot\phi X}{4H^2}\left(3h_3-\frac{h_5V}{6g}\right),
\quad
\nonumber\\&&
c_5\simeq \mu,
\bar f_i\simeq \frac{1}{2H}\zeta_{,j}h_{ij}
\quad
\bar f_{ij}\simeq -\frac{1}{4a^2H^2}\zeta_{,i}\zeta_{,j},
\end{eqnarray}
where $\mu=\dot\phi X h_5$ as defined in (\ref{mudefine}). 
It turns out that the other coefficients
are of higher order in the slow-roll parameter: $b_6\sim c_3\sim c_4\sim c_6=\mathcal{ O}(\epsilon^2)$.

\subsection{The case of k-inflation}

To extract the effect of the nontrivial sound speed,
let us consider k-inflation, which is the simplest model with
a generic value of $c_s^2$.
In the case of k-inflation, $K=K(\phi, X)$, $G_4=\mpl^2/2$, $G_3=0=G_5$,
we have
\begin{eqnarray}
\mathcal{ F}_T=\mathcal{ G}_T=\Gamma=\mpl^2,
\quad
\mathcal{ F}_S=\mpl^2\epsilon,
\quad
\mathcal{ G}_S=\frac{\mpl^2\epsilon}{c_s^2},
\quad
\Theta=\mpl^2 H, \quad\mu=0,
\end{eqnarray}
with $c_h=1$ and $r =16\epsilon c_s$, which simplifies
the coefficients in the cubic Lagrangians:
\begin{eqnarray}
&&
b_1=b_2=\frac{\mpl^2\epsilon}{8},
\quad
b_3=-\frac{\mpl^2\epsilon }{4c_s^2},
\quad
b_4=b_5=b_6=b_7=0,
\quad
\nonumber\\&&
E_{shh}=\frac{1}{4H}\zeta\dot h_{ij}E_{ij}^h,
\\
&&c_1=\mpl^2\epsilon,\quad
c_2=0,
\quad
c_3=\frac{\mpl^2\epsilon^2}{2c_s^2},
\quad
c_4=c_5=0,
\quad
c_6=\frac{\mpl^2\epsilon^2}{4c_s^4},
\label{kc}\\
&&
\bar f_i=\frac{1}{2H}\zeta_{,j}h_{ij},
\quad
\bar
f_{ij}=\frac{\epsilon}{2Hc_s^2}\zeta_{,i}\psi_{,j}-\frac{1}{4a^2H^2}\zeta_{,i}\zeta_{,j}. 
\end{eqnarray}
Note that in deriving the above coefficients
we have not invoked the slow-roll expansion.

\section{Discussion}

 In this paper we have presented the full bispectra, including the cross
bispectra of the primordial curvature and tensor perturbations, in the
generalized G-inflation model which is  the most general
single-field inflation model with the second order equations of motion.

In the event full observations of these quantities could be made, we could
extract many pieces of interesting information on the underlying theory.
For example, by observing three-point tensor correlation function, 
we can in principle determine the kinetic coupling to the Einstein
tensor through $\mu$.  Another interesting quantity is the cross
bispectrum of two tensors and one scalar.  If we could observationally
identify their coefficients $b_2,~b_3$ and $b_6$, we could in principle
determine $\mathcal{ F}_S$,  $\mathcal{ G}_S$, $\mathcal{ F}_T$, and  $\mathcal{ G}_T$
independently with the help of the three-tensor bispectrum
which would provide a consistency relation of the theory for the
tensor-to-scalar ratio (\ref{tensortoscalar}).

Let us next turn to two-scalar and one-tensor bispectrum whose 
effective Lagrangian is given by (\ref{ssh}).  Its most interesting
component is the first term proportional to $c_1=\mathcal{ F}_S$ which could be
singled out by taking $k_3$  small.  In the standard canonical inflation
as well as in k-inflation, the coefficient simply takes 
$\displaystyle{c_1=\mathcal{ F}_S=\mpl^2\epsilon=\frac{\mpl^2r}{16c_s}}$ 
as derived in (\ref{kc}), where we have used the consistency relation
in the last equality.

We can also show that this 
feature remains valid in the case where a sizable {\it local}
non-Gaussianity is generated as in the cases of the 
curvaton scenario \cite{Lyth:2002my} and the
modulated reheating scenarios \cite{Zaldarriaga:2003my}.
In such case curvature perturbation $\zeta$ is sourced by another scalar field
which we denote by $\sigma$ and its fluctuation by $\delta\sigma$.  
One can relate $\zeta$ and $\delta\sigma$ as
\begin{equation}
 \zeta=N_\sigma(\sigma)\delta\sigma + \frac{1}{2}
N_{\sigma\sigma}(\sigma)(\delta\sigma)^2, \label{deltaN}
\end{equation}
using the $\delta N$-formalism \cite{Starobinsky:1986fxa}.
Suppose that $\sigma$ has the Lagrangian $\mathcal{
L}_\sigma=\kappa(Y,\sigma)$
with $Y:=-(\partial \sigma)^2/2$.  Since the dynamics of $\sigma$ is 
practically frozen during inflation and it practically behaves as a
massless minimally-coupled field, one can expand 
$\mathcal{ L}_\sigma= \kappa(0,\sigma_0)+\kappa_\sigma(0,\sigma_0) Y$
in this regime where $\sigma_0$ is its expectation value in the domain
including  our horizon today.  Then the mean-square
fluctuation amplitude of $\sigma$ is given by 
\begin{equation}
\langle(\delta\sigma)^2\rangle=\frac{H^2}{4\pi^2 \kappa_\sigma(0,\sigma_0)}
=\frac{1}{N_\sigma^2(\sigma_0)}\mathcal{ P}_\zeta,  
\end{equation}
the latter being an outcome of (\ref{deltaN}), and it determines the
relation between $\delta\sigma$ and $\zeta$, too.  Then the effective
Lagrangian representing tensor-scalar-scalar coupling is generated from
the kinetic term of $\sigma$ in this case and reads
\begin{equation}
 \mathcal{
  L}_{ssh}=\frac{1}{2}\kappa_\sigma(0,\sigma_0)h^{\mu\nu}\sigma_{,\mu}
\sigma_{,\nu}=\frac{1}{2}\left(\frac{H}{2\pi}\right)^2\mathcal{ P}_\zeta
h^{\mu\nu}\zeta_{,\mu}\zeta_{,\nu}=\frac{M_{\rm
Pl}^2r}{16}h_{ij}\zeta_{,i}\zeta_{,j}.
\end{equation}
Note that in this case the sound speed is equal to unity.  Thus we find
that if the sector responsible for the generation of curvature perturbations
is minimally coupled to gravity with no extra Galileon-like terms, 
$c_1$ takes the same form whether they are generated
by the inflaton or another scalar field.  Thus this term can provide 
a test of the generalized Galileon as a source of the structure of the
Universe. 

It is a non-trivial issue how to normalize the cross
bispectra. In this paper, we have normalized them by the power spectrum
of the curvature perturbation. This is mainly because these cross
bispectra generate the auto- and the cross-bispectra of the temperature
fluctuation and the E-mode polarization, which are mainly sourced by the
curvature perturbation. However, such a normalization may be inadequate
for the cross bispectra including the B-mode polarization. Therefore, we
need to directly investigate the impacts on the CMB bispectra
and it is interesting to see the CMB cross-bispectra between the temperature fluctuations and B-mode polarizations which are sourced directly from the primordial cross-bispectra of the scalar and the tensor modes
\cite{Shiraishi:2010kd, Shiraishi:2011st}. Constraining  the model
parameters by CMB bispectra is a work in progress \cite{CMBbispectra}.

\section*{Acknowledgment}

~
This work was supported in part by ANR (Agence Nationale de la
Recherche) grant ``STR-COSMO'' ANR-09-BLAN-0157-01 (X.G.), the
Grant-in-Aid for JSPS Research under Grant Nos.~22-7477 (M.S.) and 24-2775 (S.Y.), the
Grant-in-Aid for Scientific Research Nos.~24740161 (T.K.), 21740187
(M.Y.) and 23340058 (J.Y.) and the Grant-in-Aid for
Scientific Research on Innovative Areas No.~24111706 (M.Y.) and 21111006
(J.Y.).


\end{document}